\documentclass[pra]{revtex4}
\usepackage{graphicx}

\begin{document}

\title{Phase transitions in the boson-fermion resonance model in one dimension}
\author{E. Orignac}
\affiliation{Laboratoire de Physique Th\'eorique de l'\'Ecole Normale
  Sup\'erieure CNRS-UMR8549 \\ 24, Rue Lhomond  F-75231 Paris Cedex 05
  France}
\author{R. Citro}
\affiliation{Dipartimento di Fisica ``E. R. Caianiello'' and
Unit{\`a} C.N.I.S.M. di Salerno\\ Universit{\`a} degli Studi di
Salerno, Via S. Allende, I-84081 Baronissi (Sa), Italy}
\date{\today}
\begin{abstract}
We study 1D fermions with photoassociation or with a narrow
Fano-Feshbach resonance described by the Boson-Fermion resonance model. Using the
bosonization technique, we derive a low-energy Hamiltonian of the
system. We show that at low energy, the order parameters for the
Bose Condensation and fermion superfluidity become identical,
while a spin gap and a gap against the formation of phase slips
are formed. As a result of these gaps, charge density wave
correlations decay exponentially in contrast with the phases where
only bosons or only fermions are present. We find a Luther-Emery
point where the phase slips and the spin excitations can be
described in terms of pseudofermions. This allows us to provide
closed form expressions of the density-density correlations and
the spectral functions. The spectral functions of the fermions are
gapped, whereas the spectral functions of the bosons remain
gapless. The application of a magnetic field results in a loss of
coherence between the bosons and the fermion and the disappearance
of the gap. Changing the detuning has no effect on the gap until
either the fermion or the boson density is reduced to zero.
Finally, we discuss the formation of a Mott insulating state in a
periodic potential. The relevance of our results for experiments
with ultracold atomic gases subject to one-dimensional confinement
is also discussed.
\end{abstract}

\maketitle

\section{Introduction}

Since the discovery of Bose-Einstein Condensation (BEC) of atoms
in optical traps, the field of ultracold atoms has experienced
tremendous developments in the recent
years.\cite{dalfovo99_bec_review} A first important step has been
the use of Fano-Feshbach
resonances\cite{fano_resonance,feshbach_resonance} to tune the
strength of atom-atom
interaction.\cite{stwalley_feshbach,tiesinga_feshbach}
Fano-Feshbach resonances take place when the energy difference
between the molecular state in the closed channel and the
threshold of the two-atom continuum in the open channel, known as
the detuning $\nu$, is zero\cite{duine_feshbach_review}. Near a
Fano-Feshbach resonance, the atom-atom scattering length possesses
a singularity. For $\nu>0$, atoms are stable, but the existence of
the virtual molecular state results in an effective attraction.
For $\nu<0$, the molecules are formed and possess a weakly
repulsive interaction.  Since the value of $\nu$ can be controlled
by an applied magnetic field, this allows to tune the sign and
strength of the atomic and molecular
interactions.\cite{inouye98_feshbach_na,roberts98_feshbach_rb,courteille98_feshbach_rb,vuletic99_feshbach_cs}
In particular, the use of Fano-Feshbach resonances has allowed the
observation of pairs of
fermionic\cite{jochim_bec,regal_bec,strecker_bec,cubizolles_bec}
or
bosonic\cite{donley_feshbach_exp,duerr04_molecules_rb,chin03_molecules_cs,xu03_molecules_bec}
atoms binding together to form bosonic molecules.  At sufficiently
low temperature, for $\nu<0$, these molecules can form a
Bose-Einstein condensate. In the case of a fermionic system, for
$\nu>0$, due to attractive interactions a BCS superfluid is
expected. Since the BEC and the BCS state break the same U(1)
symmetry, a smooth crossover between the two states is expected as
$\nu$ is tuned through the resonance.  Indeed, the BEC of
molecules\cite{jochim_bec,greiner_bec,zwierlein_bec} and the
crossover to a strongly degenerate Fermi
gas\cite{bartenstein_bec,zwierlein04_bec,regal_bec_pairs,bourdel04_bcs_bec}
have been observed as a gas of cold fermionic atoms is swept
through the Fano-Feshbach resonance. Measurement of the
radio-frequency excitation spectra\cite{chin04_gap_bcs} and of the
specific heat\cite{kinast05_Cp} as well as observation of vortices
in a rotating system\cite{zwierlein05_vortices} on the $\nu>0$
side revealed the presence of a superfluid BCS gap, thus proving
the existence of a BEC-BCS crossover.
  Such a crossover is
naturally described by the boson-fermion model, \cite{timmermans01_bosefermi_model,holland01_bosefermi_model,ohashi02_bcsbec,ohashi03_transition_feshbach,ohashi03_collective_feshbach,stajic04_bcs_bec,chen_bcs_bec_review,domanski05_feshbach}%
first introduced in the 1950s in the context of the theory of
superconductivity\cite{schafroth54_preformed_pairs,blatt64_preformed_pairs}
and later reinvestigated in the 1980s in the context of
polaronic\cite{alexandrov81} and high-Tc superconductivity
theory.\cite{ranninger85_bosefermi,friedberg89_bosefermi,geshkenbein97_preformed_pairs}
A second important parallel development has been the possibility
to form quasi-1D condensates using anisotropic
traps\cite{grimm_potential_review,hellweg01_bec1d,goerlitz01_bec1d,richard03_bec1d},
two-dimensional optical
lattices\cite{greiner01_2dlattice,moritz03_bec1d,kinoshita_tonks_experiment,paredes_toks_experiment,stoeferle_coldatoms1d,koehl_1dbose}
or atoms on chips.\cite{reichel} In one dimensional systems
interactions are known to lead to a rich
physics.\cite{giamarchi_book_1d} In particular, strongly
correlated states of fermions, where individual particles are
replaced by collective spin or density excitations, are
theoretically
expected.\cite{giamarchi_book_1d,cazalilla_1d_bec,recati03_fermi1d}
When the interactions between the fermions are repulsive, both the
spin and density fluctuations are gapless with linear dispersion
and this state is known as the Luttinger
liquid\cite{luther_bosonisation,haldane_luttinger,giamarchi_book_1d}.
For attractive interactions between the fermions, the spin degrees
of freedom develop a gap, yielding a state known as the
Luther-Emery liquid.\cite{giamarchi_book_1d,luther_exact}
Similarly, bosons are expected to be found in a Luttinger liquid
state, with individual particles being replaced by collective
density
excitations\cite{giamarchi_book_1d,cazalilla_1d_bec,haldane_bosons,petrov04_bec_review}.
Moreover, strong repulsion can lead to the fermionization of
interacting bosons i.e. the density matrix becomes identical to
that of a non-interacting spinless fermion system, the so-called
Tonks-Girardeau (TG)
regime.\cite{girardeau_bosons1d,schultz_1dbose} Experiments in
elongated traps have provided evidence for one-dimensional
fluctuations\cite{hellweg01_bec1d,goerlitz01_bec1d,richard03_bec1d}.
However, in these systems, the bosons remain weakly interacting.
With two-dimensional optical lattices, it is possible to explore a
regime with stronger repulsion. In particular, it was possible to
observe the TG regime with ${}^{87}$Rb
atoms\cite{kinoshita_tonks_experiment} by increasing the
transverse confinement. The TG regime can also be reached by
applying a 1D periodic potential along the tubes to increase the
effective mass of the bosons\cite{paredes_toks_experiment}.  Using
a stronger 1D potential, it is possible to drive a one-dimensional
Mott transition between the superfluid state and an insulating
state\cite{stoeferle_coldatoms1d}. Another characteristic of atoms
in a one-dimensional trap is that transverse confinement
 can give rise to a type of
Fano-Feshbach resonance as a function of the trapping frequency
called the confinement induced resonance
(CIR).\cite{olshanii_cir,bergeman_cir,yurovsky_feshbach} Recently,
experiments have been performed on ${}^{40}$K fermionic atoms in a
one dimensional trap forming bound states either as a result of
Fano-Feshbach resonances or of CIR.\cite{moritz05_molecules1d}
Both types of bound states have been observed and the results can
be described using the Boson-Fermion
model.\cite{dickerscheid_comment} This prompts the question of
whether a one dimensional analogue of the BEC-BCS crossover could
be observed in such a system.  It is well known that in one
dimension, no long range BEC or BCS order can
exist.\cite{mermin_wagner_theorem,mermin_theorem} However,
quasi-long range superfluid order is still possible. For fermions
with attractive interactions, it was shown using the exactly
solved Gaudin-Yang model\cite{gaudin_fermions,yang_fermions} that
for weakly attractive interactions, a Luther-Emery state with
gapless density excitations and gapful spin excitations was
formed, whereas for strongly attractive interactions the system
would crossover to a Luttinger liquid of
bosons.\cite{tokatly_bec_bcs_crossover1d,fuchs_bcs_bec} The
boson-fermion model was also considered in the case of a broad
Fano-Feshbach resonance.\cite{fuchs04_resonance_bf} In that case
only bosons or fermions are present (depending on which side of
the resonance the system is) and the results are analogous to
those obtained with the Gaudin-Yang model. In fact, in the three
dimensional case, it is possible to derive a mapping of the
boson-fermion model with a broad resonance to a model with only
fermions and a two-body interaction.\cite{simonucci05_becbcs} In
the narrow resonance case, such a mapping is valid only very close
to the resonance. It was therefore interesting to investigate what
happens in one dimension in the case of a narrow resonance.
Indeed, in the latter case, it has been shown
previously\cite{sheehy_feshbach,citro05_feshbach} that a richer
phase diagram could emerge with a phase coherence between a fluid
of atoms and a fluid of molecules at weak repulsion and a
decoupling transition for stronger repulsion.  Analogous effects
have been discussed in the context of bosonic atoms with a
Fano-Feshbach resonance in\cite{lee05_feshbach}.  Due to the
concrete possibility of forming 1D Fermi and Bose gas with optical
lattices \cite{koehl_1dbose,petrov04_bec_review} some of the
theoretical predictions in the narrow resonance case may become
testable experimentally in the future.  Experimental signature of
the phase coherence between the two fluids include density
response and momentum distribution function. In the present paper,
we investigate in more details the phase in which the atomic and
the molecular fluid coexist. In particular, we study the
equilibrium between the atomic and the molecular fluid as the
detuning is varied. Also, we investigate the effect of placing the
system in a periodic potential and show that the phase coherence
between the atomic and molecular fluid hinders the formation of
the Mott state in systems at commensurate filling. Such conclusion
is in agreement with a study in higher
dimension\cite{zhou05_mott_bosefermi}.

The plan of the paper is the following. In
Sec.\ref{sec:hamiltonian} we introduce the boson-fermion
Hamiltonian both in the lattice representation and in the
continuum. We discuss its thermodynamics in the limit of an
infinitesimal boson-fermion conversion term and show under which
conditions atoms and molecules can coexist.  In
Sec.\ref{sec:boson-appr} we derive the bosonized expression for
the boson-fermion Hamiltonian valid in the region where atoms and
molecules coexist. This Hamiltonian is valid for a system in an
optical lattice provided it is at an incommensurate filling (i.e.
with a number of atoms per site which is not integer). We show
that for not too strong repulsion in the system, a phase where the
atomic and the molecular superfluid become coherent can be
obtained. This phase possesses a spin gap. We show that in this
phase the order parameter for the BEC and the BCS superfluidity
order parameter are identical, while charge density wave
correlations present an exponential decay. We discuss the phase
transitions induced by the detuning, the magnetic field and the
repulsion. We also exhibit a solvable point where some correlation
functions can be obtained exactly.  In
Sec.~\ref{sec:mott-insul-state}, we consider the case where the
number of atoms per site in the optical lattice is integer. We
show that a phase transition to a Mott insulating state can be
obtained in that case. However, there is no density wave order in
this Mott state. Finally, in Sec.~\ref{sec:param-boson-hamilt}, we
discuss the applicability of our results to experiments.

\section{Hamiltonians and thermodynamics}
\label{sec:hamiltonian}

\subsection{Hamiltonians}
\label{sec:lattice-continuum}

 We consider a system of 1D fermionic atoms
 with a Fano-Feshbach
resonance.\cite{maccurdy_feschbach,stwalley_feshbach,tiesinga_feshbach,holland01_bosefermi_model}
 This 1D system can be obtained by trapping the fermions in a two
 dimensional or a three dimensional optical lattice. In the first
 case, the fermions are trapped into 1D tubes, in the second case, a
 periodic potential is superimposed along the direction of the tubes.
In the case in which the fermions are injected in a uniform potential,
the Hamiltonian of the system reads:
\begin{eqnarray}  \label{eq:nolattice}
H=&&-\int dx \sum_\sigma \psi^\dagger_\sigma \frac
{\nabla^2}{2m_F} \psi_\sigma + \int dx \psi_b^\dagger \left(-\frac
{\nabla^2} {2m_B} +\nu\right)\psi_b + \lambda \int dx
(\psi_b^\dagger \psi_\uparrow \psi_\downarrow +
\psi^\dagger_\downarrow \psi^\dagger_\uparrow \psi_b) \nonumber
\\ &&+ \frac 1 2
\int dx dx^{\prime}\left[V_{BB}(x-x^{\prime}) \rho_b(x)
\rho_b(x^{\prime}) +  V_{FF}(x-x^{\prime})
\sum_{\sigma,\sigma^{\prime}} \rho_\sigma(x)
\rho_{\sigma^{\prime}}(x^{\prime}) +2V_{BF}(x-x^{\prime})
\sum_{\sigma} \rho_\sigma(x) \rho_b(x^{\prime})\right],
\end{eqnarray}
\noindent where $\psi_b$ annihilates a molecule, $\psi_\sigma$ a
fermion of spin $\sigma$, $m_F$ is the mass of the isolated fermionic atom,
$m_B=2m_F$ the mass of the molecule, $V_{BB},V_{BF},V_{FF}$ are
 (respectively) the molecule-molecule, atom-molecule and atom-atom
 interactions.  Since these interactions are short ranged, it is
 convenient to assume that they are of the form
 $V_{\alpha\beta}(x)=g_{\alpha\beta} \delta(x)$. The term $\nu$ is the
 detuning.
Finally, the term $\lambda$ allows the transformation of a pair of
fermions into a  Fano-Feshbach molecule and the reverse process.
This term can be viewed as a Josephson
coupling\cite{tinkham_book_superconductors} between the order
parameter of the BEC of the molecules, and the order parameter for
the superfluidity of the fermions.  As a result of the presence of
this term, pairs of atoms are converted into molecules and
vice-versa, as in a chemical
reaction\cite{schafroth54_preformed_pairs}.  As a result of this,
only the total number of atoms (paired and unpaired),
$\mathcal{N}=2N_b+N_f$ (where $N_b$ is the number of molecules and
$N_f$ is the number of unpaired atoms) is a conserved quantity.

 In the case where atoms are injected in a periodic potential,
 $V(x)=V_0 \sin^2 (\pi x/d)$  it is convenient to introduce  the
 Wannier orbitals\cite{ziman_solid_book}  of this potential. In the
 single band approximation the Hamiltonian reads:
\cite{orso05_feshbach1d,jaksch05_coldatoms,dickerscheid_feshbach_lattice,dickerscheid_feschbach_bf}
\begin{eqnarray}  \label{eq:lattice}
H&=&-t\sum_j (f^\dagger_{j+1,\sigma} f_{j,\sigma}
+f^\dagger_{j,\sigma} f_{j+1,\sigma}) + U \sum_j n_{f,j,\uparrow}
n_{f,j,\downarrow}  \nonumber \\ && -t^{\prime}\sum_j
(b^\dagger_{j+1} b_j +b^\dagger_j b_{j+1}) +
U^{\prime}\sum_j(n_{b,j})^2 +\nu \sum_j b^\dagger_j b_j  \nonumber
\\ && + \bar{\lambda} \sum_j (b^\dagger_j
f_{j,\uparrow}f_{j,\downarrow}
+f^\dagger_{j,\uparrow}f^\dagger_{j,\downarrow} b_j ) + V_{bf} \sum_j
n_{b,j}(n_{f,j,\uparrow} + n_{f,j,\downarrow}),
\end{eqnarray}
\noindent where $f_{j,\sigma}$ annihilates a fermion of spin $\sigma$  on
site $j$, $n_{f,j,\sigma}=f^\dagger_{j,\sigma} f_{j,\sigma}$, $%
b^\dagger_{j}$ creates a Fano-Feshbach molecule (boson) on the site
$j$, and $n_{b,j}=b^\dagger_j b_j$. The hopping integrals of the
fermions and bosons are respectively $t$ and $t^{\prime}$. The
quantity $\nu$ is
the detuning.  The parameters $U$, $%
U^{\prime}$ and $V_{bf}$ measure (respectively) the
fermion-fermion, boson-boson, and fermion-boson repulsion. The
case of hard core bosons corresponds to $U'\to \infty$. The
conversion of atoms into molecules is measured by the term
$\bar{\lambda}$. Again, only the sum ${\cal
  N}=2N_b+N_f$ is conserved. We note that within the single band
approximation, there should exist a hard core repulsion between the
bosons.

\subsection{Thermodynamics of the boson-fermion model in the limit of
   $\lambda \to 0$}
 \label{sec:continuum-case}

In this Section, we wish to study the behavior of the density of
unpaired atoms $\rho_f$  and of the density of atoms paired in
molecules $\rho_b$ as a function of the total density of atoms
(pair and unpaired) $\rho_{\text{tot.}}$ in the limit of $\lambda
\to 0_+$.  In such a limit, the fermion-boson conversion does not
affect the spectrum of the system compared to the case without
fermion-boson conversion. However, it is imposing that only the
total
 total number of atoms ${\cal N}=2N_b+N_f$ is conserved. Therefore,
it this limit there is a single chemical potential $\mu$  and the partition
 function reads:
 \begin{eqnarray}\label{eq:partition-conversion}
 Z_\lambda[\mu]=\mathrm{Tr}[e^{-\beta[H_\lambda-\mu(N_f+2N_b)]}],
 \end{eqnarray}
and:
\begin{eqnarray}\label{eq:total-number}
 N_f+2N_b=\frac{1}{\beta Z_\lambda} \frac{\partial Z_\lambda}{\partial \mu},
 \end{eqnarray}
 In the absence of fermion-boson conversion,  $N_b$ and $N_f$ would be  separately conserved, and one would have a chemical potential $\mu_b$ for the molecules and $\mu_f$ for the atoms.
 The partition function of this
hypothetical system would read:
 \begin{eqnarray}\label{eq:partition-no-conversion}
 Z_0[\mu_f,\mu_b]=\mathrm{Tr}[e^{-\beta[H_0-\mu_f N_f- \mu_b N_b]}],
 \end{eqnarray}
 and thus:
 \begin{eqnarray}
 \lim_{\lambda \to 0_+} Z_\lambda[\mu]=Z_0[\mu,2\mu].
 \end{eqnarray}
If we further assume that $V_{BF}=0$, we have $%
 H_0=H_f+H_b$, where $H_f$ is the Hamiltonian of the fermion subsystem and $H_b$ is the Hamiltonian of the bosonic subsystem, and the partition function~(\ref{eq:partition-no-conversion})
 factorizes as $Z_0[\mu_f,\mu_b]=Z_f[\mu_f] Z_b[\mu_b]$, where $Z_{\nu}[\mu_\nu]=\mathrm{Tr}[e^{-\beta[H_\nu-\mu_\nu N_\nu}]$ for $\nu=f,b$. Thus,
 in the limit $\lambda,V_{BF}\to 0$, we obtain the following expression of the number of unpaired atoms $N_f$ and the number of atoms paired in molecules $N_b$.
 \begin{eqnarray}\label{eq:atom-number}
 N_f=&=&\frac{1}{\beta Z_f} \left(\frac{\partial Z_f}{\partial \mu_f}%
 \right)_{\mu_f=\mu}, \\
\label{eq:molecule-number}
 N_b&=&\frac{1}{\beta Z_b} \left(\frac{\partial Z_b}{\partial \mu_b}%
 \right)_{\mu_b=2\mu}.
 \end{eqnarray}
 We now use these equations~(\ref{eq:atom-number}) and
 (\ref{eq:molecule-number}) to study the coexistence of bosons and
 fermions as the detuning $\nu$ is varied.  Two simple cases can be
 considered to illustrate this problem of coexistence.  First, one can
 consider bosonic molecules with hard core repulsion and
 noninteracting fermionic atoms.  In such a case, the thermodynamics
 of the gas of molecules is reduced to that of a system of spinless
 fermions by the Jordan-Wigner
 transformation\cite{jordan_transformation,girardeau_bosons1d,schultz_1dbose},
 and the expression of the densities of unpaired atoms and molecules
 can be obtained in closed form. In this simple case, it is
 straightforward to show that for sufficiently negative detuning all
 atoms are paired into molecules, and for sufficiently positive
 detuning all the atoms remain unpaired.  The case of intermediate
 detuning is more interesting as coexistence of unpaired atoms with
 atoms paired into molecules becomes possible.  The physical origin of
 this coexistence is of course the molecule-molecule repulsion that
 makes the chemical potential of the gas of molecules increase with
 the density so that in a sufficiently dense gas of molecules, it
 becomes energetically favorable to create unpaired atoms.  To show
 that the above result is not an artifact of having a hard core
 repulsion, we have also considered a slightly more realistic case of
 molecules with contact repulsion and non-interacting atoms. Although
 in that case we cannot anymore obtain closed form expressions of the
 density of molecules, we can still calculate numerically the density
 of molecules using the Lieb-Liniger solution\cite{lieb_bosons_1D}.
 We will see that having a finite repulsion between the molecules
 indeed does not eliminate the regime of of coexistence.

\subsubsection{The case of bosons with hard core repulsion}

\begin{figure}[htbp]
 \centering
 \includegraphics[width=9cm]{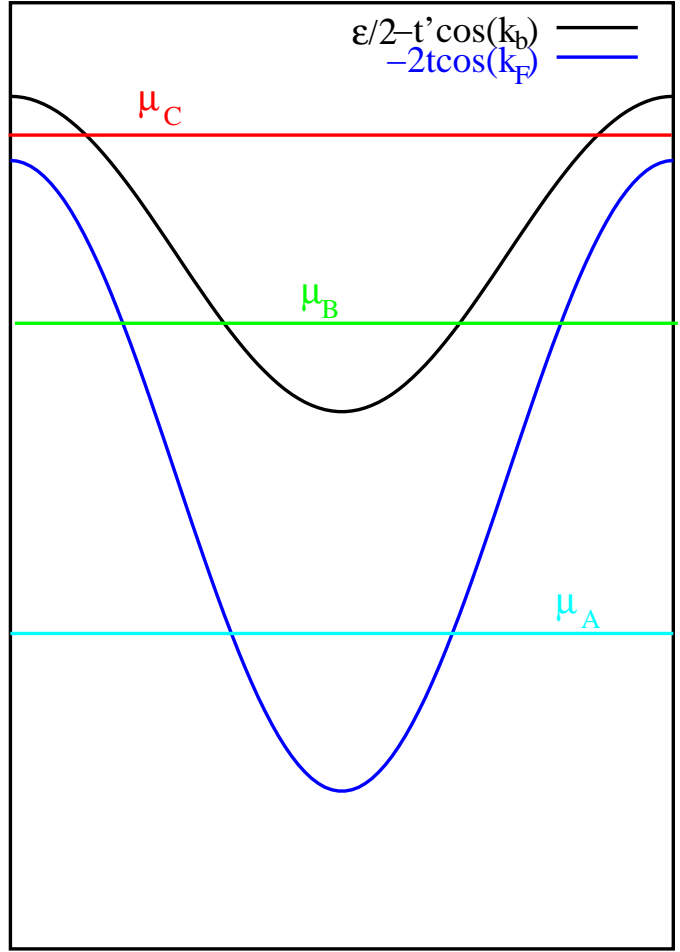}
 \caption{The different cases in the Bose-Fermi mixture with hardcore bosons
 when $\protect\nu>0$. For $\protect\mu=\protect\mu_A$, the Fermion band is
 partially filled, and the hardcore boson band is empty. For $\protect\mu=%
 \protect\mu_C$, the hardcore boson band is partially filled, and the fermion
 band is totally filled. In the case $\protect\mu=\protect\mu_B$, both band
 are partially filled. In the rest of the paper we will only consider the
 latter case.}
 \label{fig:chemical}
 \end{figure}

In that case we assume that the boson-boson repulsion $U'$ in the
lattice case and $g_{BB}$ in the continuum case is going to
infinity. Using the Jordan-Wigner
transformation\cite{jordan_transformation}, one shows that the
partition function of these hard core bosons is equal to that of
free spinless fermions thanks to the Jordan-Wigner transformation.
For positive temperature, the number of unpaired atoms and the
number of atoms paired into molecules read:
 \begin{eqnarray}\label{eq:Nf-hard-core}
 N_f&=&2 L \int \frac{dk}{2\pi} \frac 1 {e^{\beta(\epsilon_f(k)-\mu)}+1} \\
\label{eq:Nb-hard-core}
 N_b&=& L\int \frac{dk}{2\pi} \frac 1 {e^{\beta(\epsilon_b(k)+\nu -2\mu)}+1}
 \end{eqnarray}
 For $T\to 0$, these equations reduce to:
 \begin{eqnarray}\label{eq:hard-core-conditions}
 && \rho_F=\frac{N_f}{L}=\frac {2k_F}{\pi}, \nonumber \\
 && \rho_B=\frac{N_b}{L}=\frac{k_B}{\pi},\nonumber \\
 && \mu=\epsilon_F(k_F)=\frac {\nu+\epsilon_b(k_B)} 2 ,
 \end{eqnarray}
\noindent where $k_F$ is the Fermi momentum of the atoms and $k_B$
is the Fermi momentum of the spinless fermions (i.e. the
pseudo-Fermi momentum of the molecules). Up to now, we have not
specified the dispersion of the atoms and of the molecules. In the
lattice case, these dispersion are obtained from
Eq.~(\ref{eq:lattice}) as $\epsilon_f(k)=-2t \cos(k)$ and
$\epsilon_b(k)=-2t' \cos(k)$. A graphical solution of
(\ref{eq:hard-core-conditions}) is shown on
Fig.~\ref{fig:chemical} for three different values of the chemical
potential $\mu$ and $\nu>0$.  Three different regimes are
obtained. In the first one, for $\mu=\mu_A$, only unpaired atoms
are present. In the second one for $\mu=\mu_B$, unpaired atoms and
molecules coexist. In the last one, for $\mu=\mu_C$, all the
available levels of unpaired atoms are filled, and the available
levels for molecules are partially filled. As a result, the system
behaves as if only molecules were present.  This last phase is in
fact a degenerate Tonks-Girardeau gas of
molecules\cite{girardeau_bosons1d,schultz_1dbose}. In the
intermediate regime, the fermions form a two-component Luttinger
liquid\cite{cazalilla_1d_bec,recati03_fermi1d} and the bosons form
a single component Luttinger liquid.\cite{petrov04_bec_review}
Similar calculations can be performed in the case of fermions and
bosons in the continuum described by Eq.~(\ref{eq:nolattice}).
With free fermions and hard core bosons in the continuum
Eqs.~(\ref{eq:hard-core-conditions}) become:
\begin{eqnarray}\label{eq:conditions-hardcore-continuum}
  &&\frac{k_F^2}{2m_F}=\mu,\nonumber \\
  &&\frac{k_B^2}{4m_F}+\nu = 2 \mu, \nonumber \\
  &&\frac \pi 2 \rho_{\text{tot.}}= (k_F + k_B),
\end{eqnarray}
\noindent with $\rho_{\text{tot.}}=2\rho_B+\rho_F$ the total density of
atoms.
Eliminating $k_F$ in Eq.~(\ref{eq:conditions-hardcore-continuum}),
the problem is reduced to solving a second degree equation:
\begin{eqnarray}\label{eq:kb-hardcore}
  3 k_B^2 -4\pi \rho k_B +\pi^2 \rho^2 - 4 m_F \nu =0.
\end{eqnarray}

\begin{figure}[htbp]
   \centering
   \includegraphics{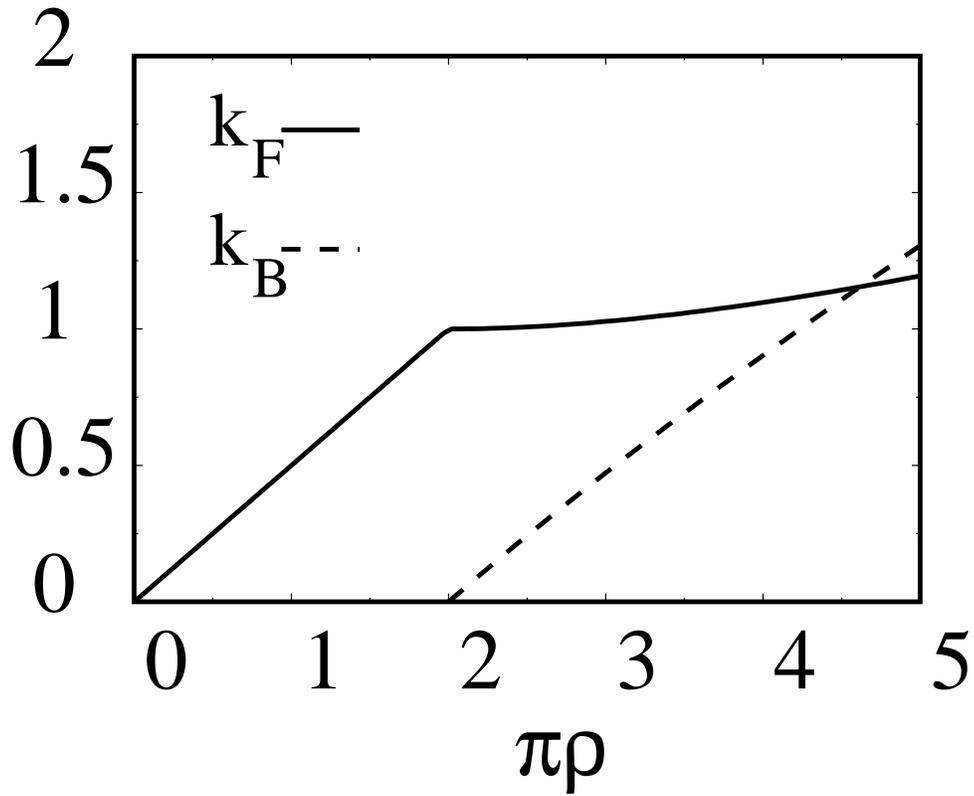}
   \caption{The behavior of $k_F$ and $k_B$ for positive detuning
   $\nu>0$ as a function of the total density $\rho$.
For low densities, only atoms are present ($k_B=0$). At
   higher densities such that $\pi\rho >2\sqrt{m_F\nu}$, a nonzero
   density of molecules  appear. At the critical density, the slopes of
   $k_F$ and $k_B$ versus $\rho$ are discontinuous. On the figure we have taken
   $m_F\nu=1$.}
   \label{fig:kf-kb1}
 \end{figure}

The solutions of Eq.~(\ref{eq:kb-hardcore}) are:
\begin{eqnarray}\label{eq:densities-hardcore}
  k_F&=&\frac 1 3 \sqrt{\pi^2 \rho^2 +12 m_F\nu} -\frac \pi 6 \rho,\nonumber \\
  k_B &=& \frac{2\pi \rho -\sqrt{\pi^2 \rho^2 + 12 m_F\nu}}{3},
\end{eqnarray}
and these solutions are physical when they yield both $k_F$ and
$k_B$ positive.  For $\nu>0$, Eq.~(\ref{eq:densities-hardcore})
yields $k_B>0$ provided
$\rho_{\text{tot.}}>\rho^{(1)}_{\text{tot.},c}=\frac 2 \pi \sqrt{m_F \nu}$.
When
$\rho<\rho^{(1)}_{\text{tot.},c}$, the density of
molecules is vanishing and $\rho_{\text{tot.}}=\rho_F$. Above the critical
density, atoms and molecules coexist, with densities given by
Eq.~(\ref{eq:conditions-hardcore-continuum}). At the critical
density, the slope of $k_B$ versus $\rho$ is discontinuous, being
$0$ below the critical density and $\frac \pi 2$ above the
critical density. The Fermi wavevector $k_F$ also possesses a
slope discontinuity at the critical density, the slope being zero
above the critical density. The behavior of $k_F$ and $k_B$ as a
function of the density is represented on Fig.~\ref{fig:kf-kb1}.

\begin{figure}[htbp]
   \centering
   \includegraphics{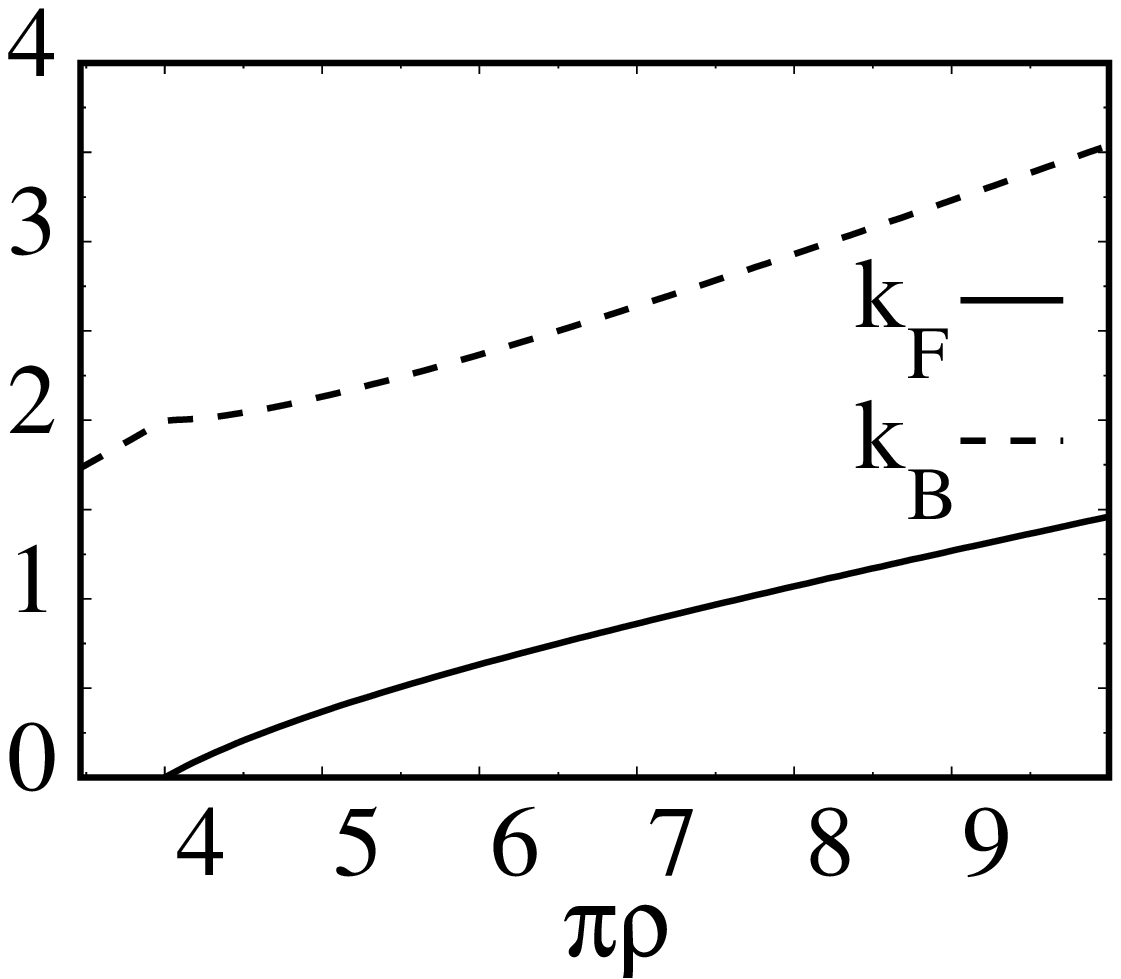}
   \caption{The behavior of $k_F$ and $k_B$ for negative detuning
   $\nu<0$. For low densities, only the molecules are present ($k_F=0$). For
   $\pi\rho>4\sqrt{m_F|\nu|}$, molecules coexist with atoms.  At the critical density, the slopes of
   $k_F$ and $k_B$ versus $\rho$ are discontinuous.   }
   \label{fig:kf-kb2}
 \end{figure}

  For $\nu<0$,
Eq.~(\ref{eq:densities-hardcore}) yields $k_F>0$ provided
$\rho>\rho^{(2)}_{\text{tot.},c}=\frac 4 \pi \sqrt{m_F|\nu|}$. When,
$\rho<\rho^{(2)}_{\text{tot.},c}$ the density of unpaired atoms vanishes,
and $\rho=\rho_B$. Above the
critical density, atoms and molecules coexist with densities given
by Eq.~(\ref{eq:conditions-hardcore-continuum}). As before, the slope of the
curve $k_F$ versus $\rho$ is discontinuous at the critical
density, being zero below and $\pi/3$ above. The behavior of $k_F$
and $k_B$ as a function of the density for $\nu<0$ is represented
on Fig.~\ref{fig:kf-kb2}.

The slope discontinuities in $k_B$ and $k_F$ have important
consequences for the compressibility. Indeed, using
Eq.~(\ref{eq:conditions-hardcore-continuum}), it is easy to see
that above the critical density, the chemical potential varies as
$O(\rho-\rho_{\text{tot.}c})^2$. Since the compressibility $\chi$
is defined as $1/\chi=\rho^2\frac {\partial \mu}{\partial \rho}$,
this implies that the compressibility of the system becomes
infinite as the critical density is approached from above,
signalling a first-order phase transition. Such first order
transitions associated with the emptying of a band have been
analyzed in the context of Luttinger liquid theory in
Refs.~\cite{nomura96_ferromagnet,cabra_instabilityLL}.

\subsubsection{The case of bosons with finite repulsion}

We have seen in the previous Section that in the case of hard core
repulsion between the molecules, both in the lattice case and in
the continuum case, that having $\nu<0$ did not prevent the
formation of unpaired atoms provided the total density of atoms
was large enough. This was related with the increase of the
chemical potential of bosons as a result of repulsion when the
density was increased. In this section, we want to analyze a
slightly more realistic case where the repulsion between bosons is
finite and check that coexistence remains possible. In the lattice
case, the problem is untractable by analytic methods and one needs
to rely on numerical
approaches.\cite{batrouni_bosons_numerique,kuhner_bosehubbard} In
the continuum case, however, it is well known that bosons with
contact repulsions are exactly solvable by Bethe Ansatz
techniques.\cite{lieb_bosons_1D} The density of molecules can
therefore be obtained by solving a set of integral
equations.\cite{lieb_bosons_1D,takahashi_tba_review} They read:
\begin{eqnarray}\label{eq:lieb_equations}
\epsilon(k)&=&\frac{\hbar^2 k^2}{2m_B} +\nu -\mu_B +\frac c \pi \int_{-q_0}^{q_0}
\frac{dq}{c^2+(q-k)^2} \epsilon(q), \\
2\pi \rho(k)&=&1+ 2c \int_{-q_0}^{q_0} \frac{\rho(q) dq}{c^2
  +(k-q)^2},
\end{eqnarray}
where:
\begin{eqnarray}
  c=\frac{m_B g_{BB}}{\hbar^2}, \\
  \rho_B=\int_{-q_0}^{q_0} \rho(q) dq,
\end{eqnarray}
$g_{BB}$ being the boson-boson interaction defined in Eq.~(\ref{eq:nolattice}).  The parameter $q_0$
plays the role of a pseudo Fermi momentum. For $q>q_0$, we have
$\rho(q)=0$. We also have $\epsilon(\pm
q_0)=0$.\cite{lieb_bosons_1D} It is convenient to introduce
dimensionless variables\cite{lieb_bosons_1D}:
\begin{eqnarray}
  \lambda=\frac c {q_0}\; ; \; \gamma = \frac c {\rho_B},
\end{eqnarray}
and rewrite $k=q_0 x$, $q=q_0 y$, $\rho(q_0 x)=g(x)$, $\epsilon(q_0
x)=\frac{\hbar^2 q_0^2}{2m} \bar{\epsilon}(x)$. The dimensionless
integral equations read:
\begin{eqnarray}\label{eq:lieb_dimensionless}
\bar{\epsilon}(x)&=& x^2+ \frac{2m(\nu -\mu_B)}{\hbar^2 q_0^2}  +\frac \lambda \pi \int_{-1}^{1}
\frac{dy}{\lambda^2+(x-y)^2} \bar{\epsilon(y)}, \\
 2\pi g(x)&=&1+ 2\lambda \int_{-1}^{1} \frac{g(y) dy}{\lambda^2
  +(x-y)^2}.
\end{eqnarray}
\noindent Using $\epsilon(\pm q_0)=0$ one has the following integral
equation for
$\bar{\epsilon}(x)$:
\begin{eqnarray}
  \bar{\epsilon}(x)= x^2 -1 +\frac \lambda \pi  \int_{-1}^{1} dy
  \bar{\epsilon}(y) \left[ \frac{1}{\lambda^2 +(x-y)^2} -
  \frac{1}{\lambda^2 +(1-y)^2}\right]
\end{eqnarray}
\noindent Once this equation has been solved, the chemical potential
  of the bosons is obtained by:
  \begin{eqnarray}
    \mu_B = \nu+ \frac{\hbar^2 q_0^2}{2m_B} \left[1+  \frac {\lambda}{\pi} \int_{-1}^{1}
    \frac{1}{\lambda^2+(x-1)^2} \bar{\epsilon}(x) dx\right].
  \end{eqnarray}

Knowing $\mu_B$ gives immediately $\mu_F=\mu_B/2$. From $\mu_F$
one finds $k_F=\sqrt{2m_F \mu_F}$ and $\rho_F=2k_F/\pi$. Finally,
using the definition of the total density $\rho=2\rho_B +\rho_F$
one can map the molecule density and the free atom density as a
function of the total density of atoms. The resulting equation of
state can be written in terms of dimensionless parameters as:
\begin{eqnarray}
  \label{eq:eq-state-adim}
  \frac{\hbar^2 \rho_B}{m_B g_{1D}} = {\cal F}\left(\frac{\hbar^2
  \rho}{m_B g_{1D}},\frac{\hbar^2 \nu}{m_F g_{1D}^2}\right)
\end{eqnarray}

The behavior of the boson density $\rho_B$ and fermion density
$\rho_F$ as a function of total density $\rho$  can be understood
in qualitative terms. Let us first discuss the case of negative
detuning. For sufficiently low densities, only bosons are present.
However, in that regime, the boson-boson repulsion is strong, and
the boson chemical potential is increasing with the boson density.
As a result, when the density exceeds a critical density $\rho_c$,
the fermion chemical potential becomes positive, and the density
of fermions becomes non-zero. The appearance of fermions is
causing a cusp in the boson density plotted versus the total
density. When the density of particles becomes higher, the
boson-boson interaction becomes weaker, and the boson chemical
potential barely increases with the density. As a result, the
fermion density becomes almost independent of the total density.
In the case of positive detuning, for low density, only fermions
are present. Again, the increase of fermion density results in an
increase of chemical potential and above a certain threshold in
fermion density, bosons start to appear, creating a cusp in the
dependence of the fermion density upon the total density. At large
density, the detuning becomes irrelevant, and the fermion density
barely increases with the total density.

\begin{figure}[htbp]
  \centering
  \includegraphics{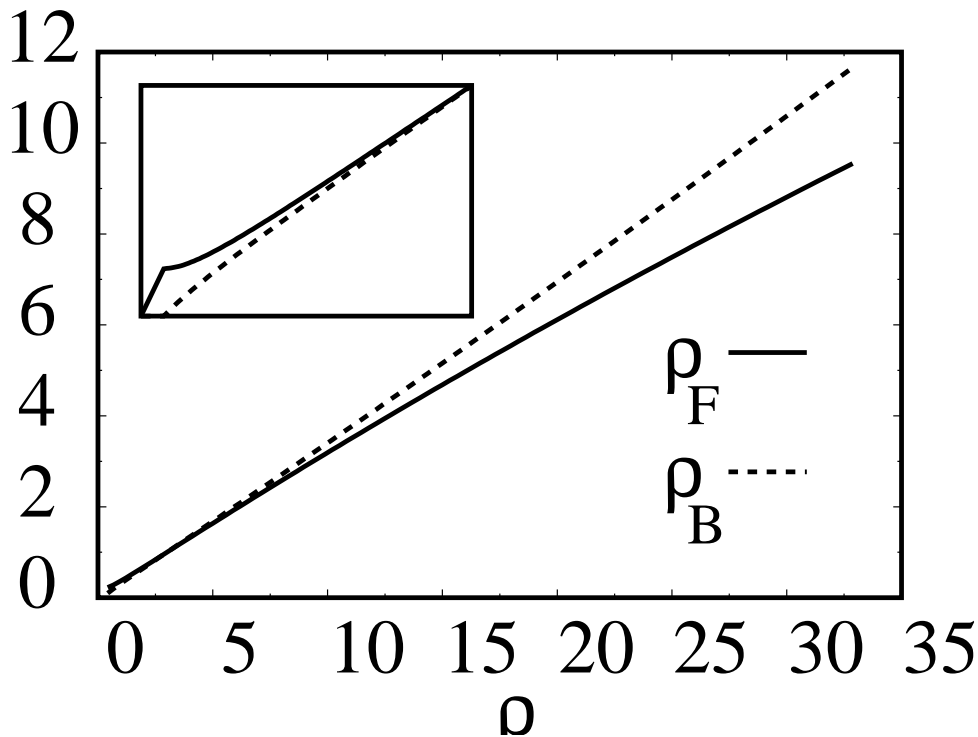}
  \caption{The density of molecules $\rho_B$ and unpaired atoms $\rho_F$ 
as a function of the
  total density $\rho$ in the case of a repulsion $c=100$  between the bosons
  and for positive detuning $\nu=0.1$. At large density, the fermion
  density is increasing more slowly than the boson density. Inset: the
  behavior of the boson and fermion densities near the origin. Note
  the cusp in the fermion density as the boson density becomes
  nonzero as in the $c=\infty$ case.}
  \label{fig:density-positive}
\end{figure}

\begin{figure}[htbp]
  \centering
  \includegraphics{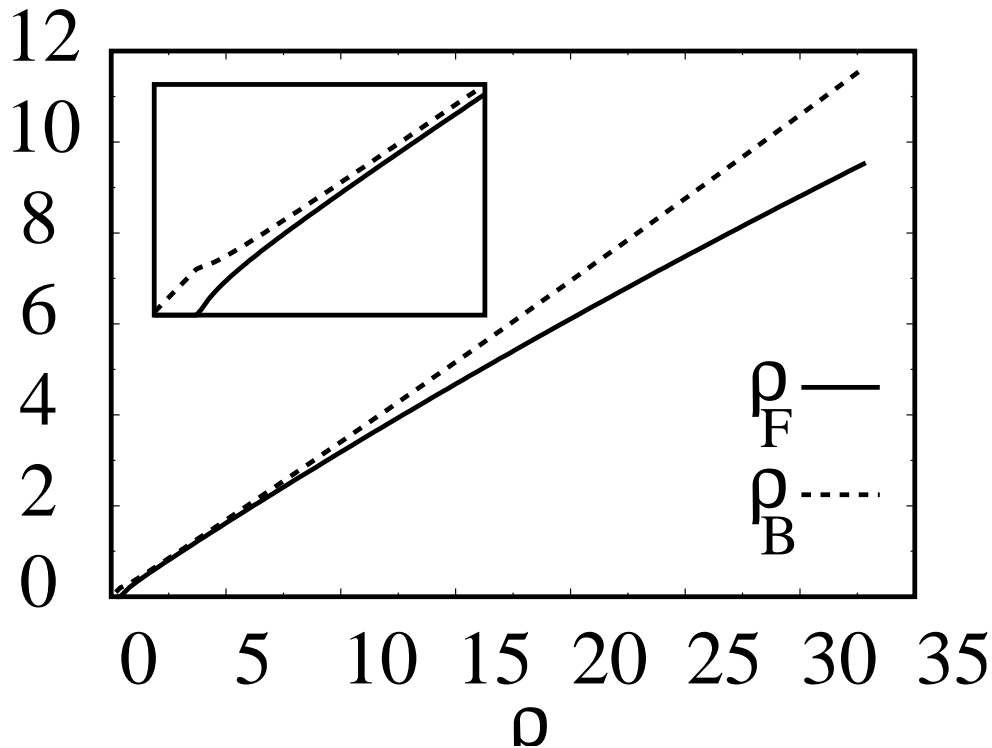}
  \caption{The density of molecules $\rho_B$ and  unpaired atoms $\rho_F$ as a function of the
  total density $\rho$ in the case of a repulsion $c=100$  between the bosons
  and for negative detuning $\nu=-0.1$. At large density, the fermion
  density is increasing more slowly than the boson density. Inset: the
  behavior of the boson and fermion densities near the origin. Note
  the cusp in the boson density as the fermion density becomes
  nonzero as in the $c=\infty$ case.}
  \label{fig:density-negative}
\end{figure}

To illustrate this behavior, we have solved numerically the integral
equations~(\ref{eq:lieb_dimensionless}), and calculated the resulting
fermion and boson densities.  A plot of the density of bosons as well
as the density of fermions is shown on Fig.~\ref{fig:density-positive}
for $\nu>0$ and on Fig.~\ref{fig:density-negative} for $\nu<0$. The
slope discontinuities at the critical density remain visible.
Obviously, this implies that the divergence of the compressibility is
still present when the repulsion between the molecule is not infinite.

We have thus seen that generally we should expect a coexistence of
fermionic atoms and bosonic molecules as soon as  repulsion between
the molecules is sufficiently strong. Moreover, the repulsion between
the molecules results in a finite velocity for sound excitations in
the molecule Bose gas. As a result, we can expect that the gas of molecules
will behave as a Luttinger liquid.  Till now however, we have
assumed that the term converting atoms into molecules was sufficiently
small not to affect significantly the spectrum of the system.
In the following, we will treat the effect of a small but not
infinitesimal conversion term  in  Eqs.~(\ref{eq:lattice}) and
(\ref{eq:nolattice}) using
  bosonization techniques. We will show that this term can lead to
  phase coherence between the atoms and the molecules, and we will
  discuss the properties of the phase in which such coherence is
  observed.

\section{Phase diagram and correlation functions}
\label{sec:boson-appr}

\subsection{Derivation of the bosonized Hamiltonian}\label{sec:deriv-boson-hamilt}

In this Section, we consider the case discussed in
Sec.~\ref{sec:hamiltonian} where neither the density of molecules
nor the density of atoms vanishes. As discussed in
Sec.~\ref{sec:hamiltonian}, this requires a sufficiently large
initial density of atoms. As there is both a non-zero density of
atoms and of molecules, they both form Luttinger
liquids\cite{petrov04_bec_review,recati03_fermi1d,cazalilla_1d_bec}.
These Luttinger liquids
 are coupled by the repulsion between atoms and molecules $V_{BF}$ and via
the conversion term or  Josephson coupling $\lambda$.
To describe these coupled Luttinger
liquids, we apply bosonization\cite{giamarchi_book_1d} to the
Hamiltonians (\ref{eq:lattice})--~(\ref {eq:nolattice}).
For the sake of definiteness,
we discuss the bosonization procedure in details only in the
case of the continuum Hamiltonian~(\ref{eq:nolattice}). For the
lattice Hamiltonian~(\ref {eq:lattice}), the steps to follow are identical
provided the system is not at a commensurate filling. At commensurate
filling, umklapp terms must be added to the bosonized Hamiltonian and
can result in Mott phases\cite{giamarchi_book_1d}. This case is treated in
Sec.~\ref{sec:mott-insul-state}.

To derive the bosonized Hamiltonian describing the low-energy spectrum of the Hamiltonian (\ref{eq:nolattice}), we need first to consider
the bosonized description of the  system when all
atom-molecule interactions are turned off. For
$\lambda=0,V_{BF}=0$, both $N_f$ and $N_b$ are conserved and the
bosonized Hamiltonian equivalent to  (\ref{eq:lattice}) or (\ref
{eq:nolattice}) is given by:
\begin{eqnarray}  \label{eq:bosonized-spin}
H&=&H_b+H_\rho+H_\sigma \nonumber \\
H_b&=&\int \frac{dx}{2\pi} \left[ u_b K_b (\pi \Pi_b)^2 +\frac {u_b}
{K_b}(\partial_x\phi_b)^2\right] \nonumber \\
H_\rho&=&\int \frac{dx}{2\pi} \left[ u_\rho K_\rho (\pi \Pi_\rho)^2 +\frac
{u_\rho} {K_\rho}(\partial_x\phi_\rho)^2\right] \nonumber \\
H_\sigma&=&\int \frac{dx}{2\pi} \left[ u_\sigma K_\sigma (\pi \Pi_\sigma)^2
+\frac {u_\sigma} {K_\sigma}(\partial_x\phi_\sigma)^2\right] -\frac{%
2g_{1\perp}}{(2\pi\alpha)^2} \int dx \cos \sqrt{8}\phi_\sigma
\end{eqnarray}
\noindent where $[\phi_{\nu}(x),\Pi_{\nu^{\prime}}(x^{\prime})]=i%
\delta(x-x^{\prime})\delta_{\nu,\nu^{\prime}}$,  ($\nu,\nu^{\prime}=b,%
\sigma,\rho$). In the context of cold atoms,  the Hamiltonian~(\ref
{eq:bosonized-spin}) have been discussed in \cite
{cazalilla_1d_bec,recati03_fermi1d,petrov04_bec_review}. The parameters $%
K_\rho$, the Luttinger exponent, and $u_\rho,u_\sigma$, the charge
and spin velocities, are known functions of the interactions
\cite{gaudin_fermions,schulz_hubbard_exact,frahm_confinv}, with
$K_\rho=1$ in the non-interacting case, $g_{1\perp}$ is a
marginally irrelevant interaction, and at the fixed point of the
RG flow $K_{\sigma}^*=1$. For the bosonic system, the parameters
$u_b,K_b$ can be obtained from numerical
calculations\cite{kuhner_bosehubbard} in the lattice case or from
the solution of the Lieb-Liniger model\cite{lieb_bosons_1D} in the
continuum case. In the case of non-interacting bosons $K_b\to
\infty$ and in the case of hard core bosons $K_b=1$.\cite{girardeau_bosons1d,schultz_1dbose,haldane_bosons} An important
property of the parameters $K_b$ and $K_\rho$ is that they
decrease as (respectively) the boson-boson and fermion-fermion
interaction become more repulsive. The bosonized Hamiltonian (\ref
{eq:bosonized-spin}) is also valid in the lattice case
(\ref{eq:lattice}) provided that both $N_f$ and $N_b$ do not
correspond to any commensurate filling.

The fermion operators can be expressed as functions of the bosonic
fields appearing in (\ref{eq:bosonized-spin})  as
\cite{giamarchi_book_1d}:
\begin{eqnarray}  \label{eq:fermion-bosonized}
\psi_\sigma(x)= \sum_{r=\pm} e^{i rk_F n \alpha}
\psi_{r,\sigma}(x=n\alpha) \\
\psi_{r,\sigma}(x) =\frac{e^{\frac{i}{\sqrt{2}} [\theta_{\rho}-r\phi_{\rho}
+\sigma (\theta_{\sigma}-r\phi_{\sigma})](x)}}{\sqrt{2\pi\alpha}},
\end{eqnarray}
\noindent where the index $r=\pm$ indicates the right/left movers,
$\alpha$ is a cutoff equal to the lattice spacing in the case of the model
Eq.~(\ref{eq:lattice}). Similarly, the  boson operators are expressed
as\cite{giamarchi_book_1d}:
\begin{eqnarray}  \label{eq:boson-bosonized}
\frac{b_{n}}{\sqrt{\alpha}} = \Psi_b(x=n\alpha) \\
\Psi_b(x)=\frac {e^{i\theta_b}}{\sqrt{2\pi\alpha}} \left[ 1 + A \cos
(2\phi_b -2 k_B x) \right].
\end{eqnarray}
\noindent In  Eqs.
(\ref{eq:fermion-bosonized})-(\ref{eq:boson-bosonized}), we have
introduced the dual fields\cite{giamarchi_book_1d} $\theta_\nu(x)
=\pi \int^x \Pi_\nu(x^{\prime})dx^{\prime}$ ($\nu=\rho,\sigma,b$),
$k_F=\pi N_f/2L$, and $k_B=\pi N_b/L$ where $L$ is the length of the
system. The fermion density is given
by\cite{giamarchi_book_1d}:
\begin{equation}  \label{eq:fermion-density-bosonized}
\sum_\sigma \frac{n_{f,n,\sigma}}{\alpha}=\rho_f(x=n\alpha)=-\frac{\sqrt{2}}{\pi}%
\partial_x\phi_\rho +\frac{\cos (2k_F x -\sqrt{2}\phi_{\rho})}{\pi \alpha}%
\cos \sqrt{2}\phi_\sigma,
\end{equation}
\noindent and the boson density by\cite{giamarchi_book_1d}:
\begin{equation}  \label{eq:boson-density-bosonized}
\frac{n_{b,n}}{a}=\rho_b(x)=-\frac{1}{\pi}\partial_x\phi_b
+\frac{\cos (2k_B x - 2\phi_b)}{\pi \alpha}.
\end{equation}

The detuning term in (\ref{eq:nolattice}) is thus expressed as:
\begin{eqnarray}
\label{eq:detun-bosonized} H_{detuning}=-\frac{\nu}{\pi}\int dx
\partial_x\phi_b
\end{eqnarray}

We now turn on a small $\lambda$ and a small $V_{BF}$.
The effect of a  small
$V_{BF}$ on a boson-fermion mixture has been investigated
previously\cite{cazalilla03_mixture,mathey04_mix_polarons}. The
forward scattering contribution is:
\begin{eqnarray}  \label{eq:V-term-bosonized}
\frac{V_{BF}\sqrt{2}} {\pi^2} \int \partial_x \phi_b \partial_x
\phi_{\rho},
\end{eqnarray}
and as discussed in \cite{cazalilla03_mixture}, it  can give rise
to a phase separation between bosons and fermions if it is too
repulsive. Otherwise, it only leads to a renormalization of the
Luttinger exponents. The atom molecule repulsion term also gives a
backscattering contribution:
\begin{eqnarray} \label{eq:cdw-locking}
  \frac{2V_{BF}}{(2\pi\alpha)^2} \int dx \cos (2\phi_b
  -\sqrt{2}\phi_\rho -2 (k_F-k_B) x)\cos \sqrt{2}\phi_\sigma,
\end{eqnarray}
however in  the general case, $k_F\ne k_B$ this contribution
is vanishing. In the special case of $k_B=k_F$, the backscattering
can result in the formation of a charge density wave.  This effect
will be discussed in Sec.~\ref{sec:quantum-ising}.
The contribution of the  $\lambda$ term is more interesting.\cite{sheehy_feshbach,citro05_feshbach}  Using Eqs.~(\ref
{eq:fermion-bosonized})-(\ref{eq:boson-bosonized}), we find that the
most relevant contribution reads:
\begin{eqnarray}  \label{eq:lambda-bosonized}
H_{bf}=\frac{2 \lambda}{\sqrt{2\pi^3 \alpha^3}} \int dx \cos (\theta_b -\sqrt{2}%
\theta_\rho) \cos \sqrt{2}\phi_{\sigma}
\end{eqnarray}

In the next section, we will see that this term gives rise to a phase
with atom-molecule coherence when the repulsion is not too strong.

\subsection{Phase diagram}
\label{sec:bf-coupling}

\subsubsection{phase with atom-molecule coherence}
The effect of the term (\ref{eq:lambda-bosonized}) on the phase
diagram can be studied by
renormalization group techniques\cite{giamarchi_book_1d}. A detailed
study of the renormalization group equations has been published in
\cite{sheehy_feshbach}. Here, we present a simplified analysis, which
is sufficient to predict the phases that can be obtained in our
system.
The scaling dimension of the boson-fermion coupling  term~(\ref
{eq:lambda-bosonized}) is: $\frac{1}{4K_b} +\frac 1 {2 K_\rho} + \frac 1 {2}
K_\sigma$. For small $\lambda$ it is reasonable  to replace $K_\sigma$ with
its fixed point value $K_\sigma^*=1$. Therefore, the RG equation for the dimensionless coupling $\tilde{\lambda}=\frac{\lambda\alpha^{1/2}}{u}$ (where $u$ is one of the velocities $u_\rho,u_\sigma,u_b$) reads:
\begin{eqnarray}\label{eq:RG-coupling}
\frac{d\tilde{\lambda}}{d\ell} = \left(\frac 3 2 -\frac 1 {2K_{\rho}} -\frac 1 {4
K_b}\right) \tilde{\lambda},
\end{eqnarray}
where $\ell$ is related to the renormalized cutoff
$\alpha(\ell)=\alpha e^{\ell}$.
We thus see that for $\frac 1 {2K_{\rho}}+\frac 1 {4 K_b} <3/2$,
this
interaction is relevant. Since for hardcore bosons\cite{girardeau_bosons1d,schultz_1dbose} $%
K_b=1$ and for non-interacting bosons $K_b=\infty$, while for free fermions $%
K_\rho=1$ and in the lattice case for $U=\infty$ one has
$K_\rho=1/2$\cite{kawakami_hubbard}, we see that the inequality is
satisfied unless  there are very strongly repulsive interactions
both in the boson system and in the fermion system. When this
inequality is not satisfied, for instance in the case of fermions
with nearest-neighbor
repulsion\cite{mila_hubbard_etendu,sano_extended_hubbard_1d}, in
which one can have $1/4<K_\rho<1/2$ and  hardcore bosons with
nearest neighbor
repulsion\cite{shankar_spinless_conductivite,haldane_luttinger},
in which one can have $K_b=1/2$, the atoms and the molecules
decouple. This case is analogous to that of the mixture of bosons and fermions\cite{mathey04_mix_polarons,cazalilla03_mixture}%
and charge density waves can be formed if $k_B$ and $k_F$ are
commensurate. The phase transition between this decoupled phase
and the coupled phase belongs to the
Berezinskii-Kosterlitz-Thouless (BKT) universality
class.\cite{kosterlitz_thouless}  As pointed out in
\cite{sheehy_feshbach}, in the decoupled phase, the effective
interaction between the fermions can be attractive. In that case,
a spin gap is formed~\cite{luther_exact,giamarchi_book_1d} and the
fermions are in a Luther-Emery liquid state with gapless density
excitations.  Let us consider the coupled phase in more details.
The relevance of the
interaction (\ref{eq:lambda-bosonized}) leads to the locking of $\phi_\sigma$%
, i.e. it results in the formation of a spin gap. To understand the effect
of the term $\cos (\theta_b  -\sqrt{2}\theta_\rho)$, it is better to perform
a rotation:
\begin{equation}
\label{eq:rot} \left(
\begin{array}{c}
\theta_{-} \\
\theta_{+}
\end{array}
\right) =\left(
\begin{array}{cc}
\frac 1 {\sqrt{3}} & -\frac{\sqrt{2}}{\sqrt{3}} \\
\frac{\sqrt{2}} {\sqrt{3}} & \frac 1{\sqrt{3}}
\end{array}
\right) \left(
\begin{array}{c}
\theta_b \\
\theta_\rho
\end{array}
\right),
\end{equation}
\noindent and the same transformation for the $\phi_\nu$.
This transformation preserves the canonical
commutation relations between $\phi_{\pm}$ and $\Pi_{\pm}$. Under this
transformation, $H_b+H_\rho$ becomes:
\begin{eqnarray}  \label{eq:b-plus-rho}
H_b+H_\rho&=&\int \frac{dx}{2\pi}\sum_{\nu=\pm} \left[ u_\nu K_\nu (\pi
\Pi_\nu)^2 + \frac{u_\nu}{K_\nu}(\partial_x \phi_\nu)^2\right]  \nonumber \\
&& + \int \frac{dx}{2\pi} [g_1 (\pi \Pi_+)(\pi \Pi_-) + g_2 \partial_x\phi_+
\partial_x\phi_-],
\end{eqnarray}
\noindent where:
\begin{eqnarray}\label{eq:corresp-uK}
u_+ K_+ &=& \frac 2 3 u_b K_b + \frac 1 3 u_\rho K_\rho, \nonumber
\\ u_- K_- &=& \frac 1 3 u_b K_b + \frac 2 3 u_\rho K_\rho,
\nonumber \\ g_1 &=& \frac{\sqrt{8}} 3 (u_b K_b -u_\rho K_\rho),
\nonumber \\ \frac{ u_+} {K_+}&=& \frac {2 u_b} {3 K_b} + \frac
{u_\rho} {3 K_\rho} + \frac{4V}{3\pi}, \\ \frac{ u_-} {K_-}&=&
\frac { u_b} {3 K_b} + \frac {2 u_\rho} {3 K_\rho} -
\frac{4V}{3\pi}, \\ g_2 &=& \frac{\sqrt{8}} 3 (\frac{u_b} {K_b}
-{u_\rho} {K_\rho} -\frac V \pi),
\end{eqnarray}
while, $H_{bf}$ defined in (\ref{eq:lambda-bosonized}%
) becomes:
\begin{eqnarray}  \label{eq:lambda-change}
H_{bf}=\frac{\lambda}{\sqrt{2\pi^3}\alpha} \int dx \cos
\sqrt{3}\theta_{-} \cos \sqrt{2}\phi_\sigma.
\end{eqnarray}
After the rotation, we see that when $\lambda$ is relevant, the field $%
\theta_-$ is also locked, but $\phi_+$ remains gapless. Since the
field $\theta_-$ is the difference of the superfluid phase of the
atoms and the one of the molecules, this means that when $\lambda$
becomes relevant, unpaired atoms and molecules share the same
superfluid phase i.e. they become coherent.  The gap induced by the
term $\lambda$ can be estimated from the renormalization group
equation~(\ref{eq:RG-coupling}). Under the renormalization group,
the dimensionless parameter $\tilde{\lambda}(\ell)$ grows until it
becomes of order one at a scale $\ell=\ell^*$ where the perturbative
 approach breaks down. Beyond the scale $\ell^*$,
the fields $\theta_-$ and $\phi_\sigma$  behave as classical fields.
Therefore, the associated energy scale $u/(\alpha e^{\ell^*})$ is
the scale of the gap. From this argument, we obtain that the gap behaves as:
\begin{equation}\label{eq:gap-RG}
\Delta \sim \frac {u}{\alpha} \left(\frac{\lambda \alpha^{1/2}}{u}\right)^{\frac 1 {\frac 3 2 -\frac 1 {2 K_\rho} -\frac {1}{4 K_b}}}.
\end{equation}
The  gapful excitations have a dispersion $\epsilon(k)=\sqrt{(uk)^2+\Delta^2}$  and are kinks and antikinks of the fields $\theta_-$ and $\phi_\sigma$.\cite{rajaraman_instanton}  More
precisely,  since a kink must interpolate between degenerate classical ground states of the potential~(\ref{eq:lambda-change}), we find that when a kink is present  $\theta_-(+\infty)-\theta_-(-\infty)=\pm
\pi/\sqrt{3}$ and $\phi_{\sigma}(+\infty)
-\phi_{\sigma}(-\infty)=\pi/\sqrt{2}$.
This indicates that a
kink is carrying a spin $1/2$, and
is making the phase $\theta_b$ of the bosons jump by $%
\pi/3$ and the phase of the superfluid order parameter $\sqrt{2}%
\theta_\rho$ of the  fermions jump by $-2\pi/3$. Since the current
of bosons is $j_b = u_b K_b \pi \Pi_b = u_b K_b \partial_x \theta_b$ and the
current of fermions is $j_F=\sqrt{2} u_\rho K_\rho \pi \Pi_\rho=\sqrt{2} u_\rho K_\rho \partial_x \theta_\rho$, this indicates that counterpropagating supercurrents of atoms and molecules exist in the vicinity of the kinks.
 Therefore, we can view  the kinks and antikinks
 are composite objects formed of vortices bound with a spin 1/2.
We
note that the kinks and antikinks may not exhaust all the possible gapful
excitations of the system. In particular, bound states of kinks
and antikinks, known as breathers
may also be present\cite{rajaraman_instanton}.
However, these gapful excitations present a larger gap than the
single kinks.
Let us now turn to the gapless
field $\phi_+$. This field  has a  simple physical interpretation.
Considering  the integral
\begin{eqnarray}  \label{eq:connection-to-N}
-\frac 1 \pi \int_{-\infty}^{\infty} dx \partial_x \phi_+ &=& -\frac 1 {\pi%
\sqrt{3}} \int_{-\infty}^{\infty} (\sqrt{2}\phi_b
+\phi_\rho)=\frac{\mathcal{N}}{\sqrt{6}},
\end{eqnarray}
 showing that $\phi_+(\infty)-\phi_+(-\infty)$  measures the
total number of particles in the system $\mathcal{N}$.
Thus $(\Pi_+,\phi_+)$ describe the total density excitations of the system.

The resulting low-energy Hamiltonian describing the gapless total
density modes reads:
\begin{eqnarray}  \label{eq:gapless-modes}
H_+=\int \frac{dx}{2\pi} \left[ u^*_+ K^*_+ (\pi \Pi_+)^2 +\frac {u^*_+}
{K^*_+}(\partial_x\phi_+)^2\right],
\end{eqnarray}
where $u^*_+,K^*_+$ denote renormalized values of $u_+,K_+$. This
renormalization is caused by the residual interactions between
gapless modes  and the gapped modes measured by  $%
g_1,g_2$ in Eq.~(\ref{eq:b-plus-rho}). Since $\phi_+$ measures the
total density, the Hamiltonian (\ref{eq:gapless-modes}) describes the
propagation of sound modes in the 1D fluid with dispersion $\omega(k)=u|k|$.
We note than in Ref.\cite{zhou05_mott_bosefermi,zhou05_mott_bosefermi_long},
dispersion relation similar to ours were derived for the sound modes and
the superfluid phase difference modes using different methods.

\subsubsection{effect of the detuning and applied magnetic field}
Having understood the nature of the ground state and the low excited
 states when $\lambda$ is relevant we  turn to the effect of the detuning term.
 Eqs.~(\ref{eq:detun-bosonized}) and~(\ref{eq:rot}) show that the
 detuning term can be expressed as a function of $\phi_+,\phi_-$ as:
\begin{eqnarray}\label{eq:detun-pm}
H_{detun.}=-\frac{\nu}{\pi} \int \partial_x \left(\sqrt{\frac 2 3}\phi_+ +
\frac{\phi_-}{\sqrt{3}} \right).
\end{eqnarray}

\begin{figure}[htbp]
\centering
\includegraphics[width=9cm]{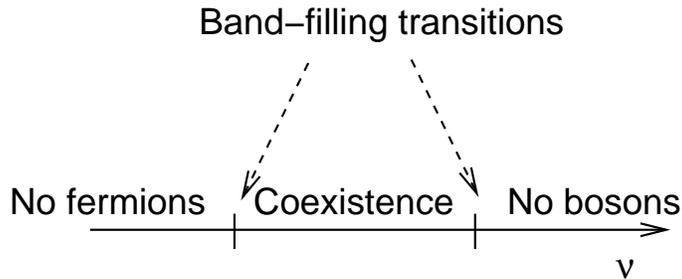}
\caption{The band filling transitions as a function of the detuning}
\label{fig:band-filling}
\end{figure}

This shows that the detuning does not affect the boson-fermion coupling~(%
\ref{eq:lambda-bosonized}) since it can be eliminated from the
Hamiltonian by a canonical transformation $\phi_\pm \to \phi_\pm
+\lambda_{\pm} x$, where $\lambda_+=\nu \sqrt{\frac 2 3}$ and
$\lambda_-=\nu \frac{1}{\sqrt{3}}$. For a fixed total density, changing
the detuning only modifies the wavevectors $k_B$ and $k_F$. As
discussed extensively in Sec.~\ref{sec:continuum-case}, for a
detuning sufficiently large in absolute value, only molecules or
only atoms are present, and near the critical value of the
detuning, the compressibility of the system is divergent. We
therefore conclude that in one-dimension, the crossover from the
Bose condensation to the superfluid state, as the detuning is
varied, is the results of the band-filling transitions at which
either the density of the atoms or the molecules goes to zero. At
such band filling transitions,$v_{\rho,\sigma}\to 0$ (respectively
$v_b\to 0$) and bosonization breaks down
\cite{nomura96_ferromagnet,cabra_instabilityLL,yang04_ferromagnet}.
The two band filling transitions are represented on Fig.~\ref
{fig:band-filling}. The cases at the extreme left and the extreme
right of the phase diagram have been analyzed in
\cite{fuchs04_resonance_bf}, where it was shown that in the case
of a broad Fano-Feshbach resonance, the zone of coexistence was
very narrow. In the narrow Fano-Feshbach resonance case we are
investigating, the zone of coexistence can be quite important.

Application of a magnetic field can also induce some phase
transitions. The interaction with the magnetic field reads:
\begin{eqnarray}  \label{eq:magfield}
H_{magn}=-\frac{h}{\pi \sqrt{2}} \int dx \partial_x\phi_\sigma,
\end{eqnarray}
The effect of the magnetic field is to
lower the gap for the creation of kink excitations (remember that they carry
a spin $1/2$). As a result, when it becomes larger than the gap,
the magnetic field induces a commensurate incommensurate transition\cite
{japaridze_cic_transition,pokrovsky_talapov_prl,schulz_cic2d,horowitz_renormalization_incommensurable}
that destroys the coherence between atoms and molecules and gives back
decoupled Luttinger liquids\cite{chitra_spinchains_field}. In that regime, the
behavior of the system is described by the models of Ref.~%
\onlinecite{cazalilla03_mixture,mathey04_mix_polarons}.
Commensurate-incommensurate transitions have been already
discussed in the context of cold atoms in \cite
{buechler03_cic_coldatoms}. In the problem we are considering,
however , since two fields are becoming gapless at the same time,
$\theta_-$ and $\phi_\sigma$ , there are some
differences\cite{orignac_spintube,yu_spin_orbital} with the
standard case\cite{buechler03_cic_coldatoms}, in particular the
exponents at the transition are non-universal.

To conclude this section, we notice that we have found three types
of phase transitions in the system we are considering. We can have
Kosterlitz-Thouless phase transitions as a function of
interactions, where we go from a phase with locked superfluid
phases between the bosons and the fermions at weak repulsion to a
phase with decoupled bosons and fermions at strong repulsion. We
can have band-filling transitions  as a function of the detuning
between the phase in which atoms and molecule coexist and phases
where only atoms or only molecules are present. Finally, we can
have commensurate-incommensurate transitions as a function of the
strength of the magnetic field. In the following section, we
discuss the correlation functions of superfluid and charge density
wave order parameters in the phase in which molecules and atoms
coexist with their relative superfluid phase $\theta_-$ locked.

\subsubsection{Quantum Ising phase transition for $k_F=k_B$}
\label{sec:quantum-ising} In the case of $k_F=k_B$, the
backscattering term (\ref{eq:cdw-locking}) is non-vanishing. This
term induces a mutual locking of the densities of the bosons and
the fermions\cite{cazalilla03_mixture} and favors charge density
wave fluctuations. This term is competing  with the Josephson
term~(\ref{eq:lambda-bosonized}) which tends to reduce density
wave fluctuations.  For $k_F=k_B$ the relevant part of the
Hamiltonian given by the combination of the terms
(\ref{eq:cdw-locking}) and (\ref{eq:lambda-bosonized}) reads:
\begin{eqnarray}\label{eq:competing}
  H_{\text{Josephon}+\text{CDW Lock.}} =\int dx \left[\frac{2\lambda}{\sqrt{2\pi^3 \alpha^3}} \cos
  ( \theta_b -\sqrt{2}\theta_\rho) +\frac{2V_{BF}}{(2\pi \alpha)^2} \cos (2\phi_b
  -\sqrt{2}\phi_\rho)\right] \cos \sqrt{2}\phi_\sigma
\end{eqnarray}

Using a transformation $\phi_b =\tilde{\phi}_b/\sqrt{2}$,
$\theta_b =\tilde{\theta}_b \sqrt{2}$, and introducing the linear
combinations
\begin{eqnarray}
  \phi_1 =\frac{\tilde{\phi}_b +\phi_\rho}{\sqrt{2}} \\
  \phi_2 = \frac{\tilde{\phi}_b - \phi_\rho}{\sqrt{2}}
\end{eqnarray}
and similar combinations for the dual fields, we can rewrite the
interaction term~(\ref{eq:competing}) as:
\begin{eqnarray}\label{eq:quantum-ising}
 H_{\text{Josephon}+\text{CDW Lock.}} =\int dx \left[\frac{2\lambda}{\sqrt{2\pi^3 \alpha^3}} \cos
 2\theta_2 +  \frac{2V_{BF}}{(2\pi \alpha)^2} \cos
 2\phi_2 \right] \cos \sqrt{2}\phi_\sigma
\end{eqnarray}
\noindent  From this Hamiltonian, it is immediate to see that a
quantum Ising phase transition occurs between the density wave
phase $\phi_2$ and the superfluid phase $\theta_2$ at a critical
point $\lambda_c=\frac{V_{BF}}{\sqrt{8\pi
\alpha}}$.\cite{finkelstein_2ch,schulz_2chains,fabrizio_dsg}
Indeed, the field $\phi_\sigma$ being locked, we can replace $\cos
\sqrt{2}\phi_\sigma$ by its expectation value in
Eq.~(\ref{eq:quantum-ising}), and rewrite (\ref{eq:quantum-ising})
as a free massive Majorana fermions
Hamiltonian.\cite{finkelstein_2ch,schulz_2chains,fabrizio_dsg} At
the point $\lambda_c$, the mass of one of these Majorana fermions
vanishes giving a quantum critical point in the Ising universality
class\cite{sachdev_book}.
 On one side of the transition, when
$\lambda>\lambda_c$, the system is in the superfluid state
discussed in Sec.~\ref{sec:bf-coupling}, on the other side
$\lambda<\lambda_c$,  the charge density wave state discussed in
\cite{cazalilla03_mixture} is recovered.

\subsection{Correlation functions}

In order to better characterize the phase in which $\lambda$ is
relevant, we need to study the correlation function of the superfluid
and the charge density wave operators. Let us begin by characterizing
the superfluid order parameters. First, let us consider the order
parameter for BEC of the molecules.
As a result of the locking of the fields $\theta_-$ and $\phi_\sigma$, the
boson operator Eq.~(\ref{eq:boson-bosonized})  becomes at low energy:
\begin{equation}
\Psi_B(x)\sim \frac{e^{i\sqrt{\frac 2 3} \theta_+}}{\sqrt{2\pi\alpha}}\left\langle e^{-i\frac{\theta_-}{\sqrt{3}}}\right\rangle.
\end{equation}
An order of magnitude of $\langle
e^{-i\frac{\theta_-}{\sqrt{3}}}\rangle$ can be obtained from a
scaling argument similar to the one giving the gap. Since the
scaling dimension of the field $ e^{-i\frac{\theta_-}{\sqrt{3}}}$
is $1/12K_{-}$, and since the only lengthscale in the problem is
$e^{\ell^*}$, we must have $ \langle
e^{-i\frac{\theta_-}{\sqrt{3}}}\rangle \sim e^{- \ell^*/12K_{-}}
\sim (\lambda \alpha^{1/2}/u)^{1/12K_{-}}$. \noindent Similarly,
the order parameter for s-wave superconductivity of the atoms
$O_{SS} = \sum_\sigma \psi_{r,\sigma} \psi_{-r,-\sigma}$ becomes:
\begin{equation}
O_{SS} = \frac{e^{i\sqrt{2}\theta_\rho}}{\pi \alpha} \cos \sqrt{2}
\phi_\sigma \sim \frac{e^{i\sqrt{\frac 2 3} \theta_+}}{\pi \alpha} \left\langle e^{\frac{2i}{\sqrt{3}}\theta_-} \cos \sqrt{2}\phi_\sigma \right\rangle,
\end{equation}
thus indicating that the order parameters of the BEC
and the BCS superfluidity have become identical in the low energy
limit\cite{ohashi03_transition_feshbach,ohashi03_collective_feshbach}.
This is the signature of the coherence between the atom
and the molecular superfluids. The boson correlator behaves as:
\begin{eqnarray}
\langle \Psi_B(x,\tau) \Psi_B(0,0)=\left(\frac{\alpha^2}{x^2 +(u \tau)^2}%
\right)^{\frac 1 {6K_+}}
\end{eqnarray}
As a result, the molecule momentum distribution becomes
$n_B(k)\sim |k|^{1/(3K_+)-1}$. One can see that the tendency
towards superfluidity is strongly enhanced since the divergence of
$n_B(k)$ for $k\to 0$ is increased by the coherence between the
molecules and the atoms. This boson momentum distribution can, in
principle be measured in a condensate expansion experiment\cite
{gerbier05_phase_mott_coldatoms,altman04_exploding_condensates}.

\begin{figure}[htbp]
  \centering
  \includegraphics[width=9cm]{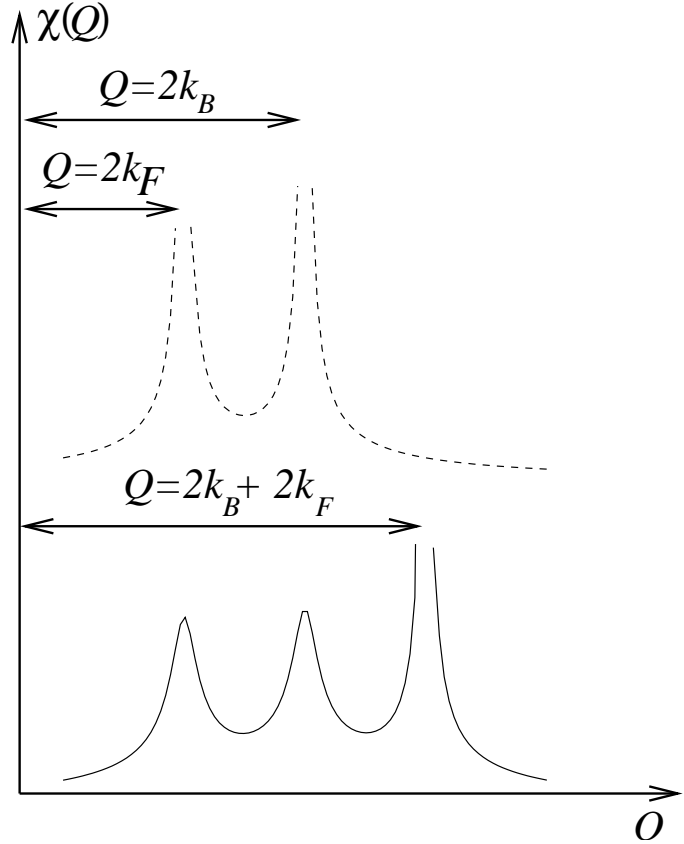}
  \caption{Fourier transform of the static density density correlations. 
In the decoupled phase (dashed line), two peaks are obtained at twice the Fermi wavevector of the unpaired atoms and at twice the pseudo-Fermi wavevector of
the molecules. In the coupled phase (solid line), the peaks are replaced by maxima at $Q=2k_F$ and $Q=2k_B$. A new peak at $Q=2(k_F+k_B)$ is obtained as
a result of Boson-Fermion coherence. }
  \label{fig:chi-q}
\end{figure}

Having seen that superfluidity is enhanced in the system, with BEC and
BCS order parameters becoming identical, let us turn to the density
wave order parameters. These order parameters are simply the staggered
components of the atom and molecule  density in
Eqs.~(\ref{eq:fermion-density-bosonized})--(\ref{eq:boson-density-bosonized}).
In terms
of $\phi_\pm$,  the staggered component of the molecule density is reexpressed  as:
\begin{eqnarray}  \label{eq:boson-staggered-gap}
\rho_{2k_B,b}(x)\sim \cos \left[2\left(\frac{\phi_-}{\sqrt{3}} +\frac{\sqrt{2%
}}{\sqrt{3}} \phi_+\right) -2k_b x\right],
\end{eqnarray}
and the staggered component of the fermion density as:
\begin{eqnarray}  \label{eq:fermion-staggered-gap}
\rho_{2k_F,f}(x)\sim \cos \left[\sqrt{2} \left(-\frac{\sqrt{2}}{\sqrt{3}}%
\phi_- +\frac{1}{\sqrt{3}} \phi_+\right) -2k_F x\right],
\end{eqnarray}
\noindent where we have taken into account the long range ordering of
$\phi_\sigma$. We see that the correlations of both $\rho_{2k_B,b}(x)$ and
$\rho_{2k_F,f}(x)$  decay exponentially due to the presence of the
disorder  field $%
\phi_-$ dual to $\theta_-$.  In more physical terms, the
exponential decay of the density-wave correlations in the system
results from the constant conversion of molecules into atoms and
the reciprocal process which prevents the buildup of a well
defined atomic Fermi-surface or molecule pseudo-Fermi surface. The
exponential decay of the density wave correlations in this system
must be be contrasted with the power-law decay of these
correlations in a system with only bosons or in a system of
fermions with attractive interactions.\cite{giamarchi_book_1d} In
fact, if we consider that our new particles are created by the
operator $\psi_b\sim \psi_\uparrow \psi_\downarrow$, we can derive
an expression of the density operators of these new particles by
considering the product $\psi_b \psi_\uparrow \psi_\downarrow$.
Using the Haldane expansion of the boson creation and annihilation
operators\cite{haldane_bosons,lukyanov_xxz_asymptotics}, we can
write this product as:
\begin{eqnarray}
 \psi^\dagger_b \psi_\uparrow \psi_\downarrow &\sim& e^{-i\theta_b} \left[\sum_{m=0}^\infty \cos (2 m \phi - 2 m k_B x) \right] \times  e^{i\sqrt{2} \theta_\rho}\left[\cos \sqrt{2}\phi_\sigma + \cos (\sqrt{2} \phi_\rho - 2 k_F x)\right] \nonumber  \\
&\sim& \langle e^{-i(\theta_b-\sqrt{2}\theta_\rho)}  \cos \sqrt{2}
\phi_\sigma \rangle \cos (\sqrt{6} \phi_+ -2 (k_F+k_B) x),
\end{eqnarray}
where $(k_F+k_B)=\pi (2N_b+N_f)/2L=\pi \rho_{pairs}$ can be
interpreted as the pseudo Fermi wavevector of composite bosons. A
scaling argument shows that the prefactor in the expression varies
as a power of $\lambda$. As a result, when there is coherence
between atoms and molecule, power-law correlations appear in the
density-density correlator near the wavevector $2k_F+2k_B$ and the
intensity of these correlations is proportional to the $ |\langle
e^{-i(\theta_b-\sqrt{2}\theta_\rho)} \cos \sqrt{2} \phi_\sigma
\rangle|^2$. The resulting behavior of the Fourier transform of
the density-density correlator is represented on Fig.~\ref{fig:chi-q}. 

 Another interesting consequence of the existence of
atom/molecule coherence is the possibility of having non-vanishing
cross-correlations of the atom and the molecule density.  In the
three-dimensional case such cross correlations have been studied
in \cite{ohashi05_bcs_bec_collective}. If we first consider
cross-correlations $\langle T_\tau \rho_{2k_B,b}(x,\tau)
\rho_{2k_F,f}(0,0)\rangle$ we notice that due to the presence of
different exponentials of $\phi_+$ in Eqs.~(\ref
{eq:boson-staggered-gap})--~(\ref{eq:fermion-staggered-gap}), this
correlator vanishes exactly. Therefore, no cross correlation
exists between the staggered densities.  However, if we consider
the cross correlations of the uniform
densities, we note that since they can all be  expressed as functions of $%
\partial_x\phi_+,\partial_x\phi_-$, such cross correlations will be
non-vanishing. More precisely, since:
\begin{eqnarray}
 \rho_F &=& -\frac{\sqrt{2}}{\pi\sqrt{3}}\partial_x\phi_+
 +\frac{2}{\pi\sqrt{3}}\partial_x\phi_- \\
 \rho_B &=& -\frac{\sqrt{2}}{\pi\sqrt{3}}\partial_x\phi_+
 -\frac{1}{\pi\sqrt{3}}\partial_x\phi_-,
\end{eqnarray}
 at low energy we have: $\rho_F \sim \rho_B \sim
 -\frac{\sqrt{2}}{\pi\sqrt{3}} \partial_x\phi_+$.

\subsection{The Luther Emery point}
\label{sec:luther_emery} In this Section, we will obtain detailed
expressions for these correlation functions at a special exactly
solvable point of the parameter space. At this point, the kinks of
the fields $\phi_\sigma$ and $\theta_-$ become free massive
fermions.
 This property, and the
equivalence of free massive fermions in 1D with the 2D non-critical Ising
model\cite{luther_ising,zuber_77,schroer_ising,kadanoff_gaussian_model,ogilvie_ising,boyanovsky_ising,itzykson-drouffe-1}
allows one to find exactly the correlation functions.

\subsubsection{mapping on free fermions}

As we have seen, after the rotation (\ref{eq:rot}), if
we neglect the interaction terms of the form $\Pi_+\Pi_-$ or $%
\partial_x\phi_+\partial_x\phi_-$, the Hamiltonian of the
massive modes $\phi_-,\phi_\sigma$ can be rewritten as:
\begin{eqnarray}  \label{eq:massive}
H&=&\int \frac{dx}{2\pi} \left[ u_\sigma^* K_\sigma^* (\pi \Pi_\sigma)^2
+\frac {u_\sigma^*} {K_\sigma^*}(\partial_x\phi_\sigma)^2\right]  \nonumber
\\
&& + \int \frac{dx}{2\pi} \left[ u_{-} K_{-} (\pi \Pi_{-})^2 +\frac {u_{-}}
{K_{-}}(\partial_x\phi_{-})^2\right]  \nonumber \\
&& + \frac{\lambda}{\sqrt{2\pi^3\alpha^3}} \int dx \cos \sqrt{3}\theta_- \cos
\sqrt{2}\phi_\sigma,
\end{eqnarray}
where $K_\sigma^*=1$. When the Luttinger exponent is $K_-=3/2$, it
is convenient to introduce the fields:
\begin{eqnarray}
\overline{\phi}=\sqrt{\frac 3 2}\theta_- \\
\overline{\theta}=\sqrt{\frac 2 3}\phi_-,
\end{eqnarray}
and rewrite the Hamiltonian~(\ref{eq:massive}) as:
\begin{eqnarray}  \label{eq:massive-bar}
H&=&\int \frac{dx}{2\pi} \left[ u_\sigma^* (\pi \Pi_\sigma)^2 +{u_\sigma^*}
(\partial_x\phi_\sigma)^2\right]  \nonumber \\
&& + \int \frac{dx}{2\pi} \left[ u_{-} (\pi \overline{\Pi})^2 + {u_{-}}
(\partial_x\overline{\phi})^2\right]  \nonumber \\
&& + \frac{\lambda}{\sqrt{2\pi^3}\alpha} \int dx \cos \sqrt{2} \overline{\phi%
} \cos \sqrt{2}\phi_\sigma.
\end{eqnarray}
If we neglect the velocity difference, i.e. assume that
$u_\sigma^*=u_-=u$, and introduce the pseudofermion fields:
\begin{eqnarray}  \label{eq:pseudofermions}
\Psi_{r,\sigma}=\frac {e^{i[(\overline{\theta}-r \overline{\phi}) +
\sigma(\theta_\sigma-r \phi_\sigma)]}}{\sqrt{2\pi\alpha}},
\end{eqnarray}
we see immediately that the Hamiltonian (\ref{eq:massive-bar}) is the
bosonized form of the following free fermion Hamiltonian:
\begin{eqnarray}  \label{eq:fermionized-ham}
H= \sum_\sigma \int dx \left[ -i u \sum_{r=\pm} r
\Psi_{r,\sigma}^\dagger \partial_x \Psi_{r,\sigma} + \frac{\lambda}{\sqrt{%
2\pi\alpha}} \Psi_{r,\sigma}^\dagger \Psi_{r,\sigma}\right].
\end{eqnarray}
\noindent As a result, for the special value of $K_-=3/2$, the
excitations can be described as massive free fermions with dispersion
$\epsilon(k)=\sqrt{(u k)^2 +m^2}$, where
$m=\frac{|\lambda|}{\sqrt{2\pi\alpha}}$. This is known as Luther-Emery
solution.\cite{luther_exact,coleman_equivalence} One can see that the
fermions carry a spin $1/2$ and a jump of the phase $\theta_-$ equal
to $\frac{\pi}{\sqrt{3}}$. Therefore they can be identified to the
kinks obtained in the semiclassical treatment of
Sec.~\ref{sec:bf-coupling}. Also, making all velocities equal and
$V_{BF}=0$ in (\ref{eq:corresp-uK}), we find the relation
$3/K_-=1/K_b+2/K_\rho$ and thus the gap given by the RG varies as
$\Delta \sim \frac{u}{\alpha} \left(\frac{\lambda
    \alpha^{1/2}}{u}\right)^{\frac 1 {\frac 3 2 - \frac{3}{4K_-}}}$.
For $K_-=3/2$ this expression reduces to the one given by the fermion
mapping.

\subsubsection{Correlation functions}

To calculate the correlation functions it is convenient to
introduce the fields:
\begin{eqnarray}
\Phi_\sigma &=& \frac 1 {\sqrt{2}} (\overline{\phi}+\sigma\phi_\sigma) \\
\Theta_\sigma &=& \frac 1 {\sqrt{2}} (\overline{\theta}+\sigma\theta_\sigma)
\end{eqnarray}
And reexpress operators by:
\begin{eqnarray}
e^{i\frac 2 {\sqrt{3}} \phi_-} &=& e^{i(\Theta_\uparrow + \Theta_\downarrow)}
\nonumber \\
e^{i\sqrt{2} \phi_\sigma} e^{- i\frac 2 {\sqrt{3}} \phi_-} &=&
e^{i(\Phi_\uparrow - \Phi_\downarrow)}e^{-i(\Theta_\uparrow +
\Theta_\downarrow)} \sim \Psi^\dagger_{+,\uparrow}
\Psi^\dagger_{-,\downarrow}  \nonumber \\
e^{-i\sqrt{2} \phi_\sigma} e^{- i\frac 2 {\sqrt{3}} \phi_-} &=&
e^{-i(\Phi_\uparrow - \Phi_\downarrow)}e^{-i(\Theta_\uparrow +
\Theta_\downarrow)} \sim \Psi^\dagger_{-,\uparrow}
\Psi^\dagger_{+,\downarrow}
\end{eqnarray}

This leads us to the following expression of the fermion density
$\rho_{2k_F,f}(x)$:
\begin{eqnarray}
\label{eq:rho} \rho_{2k_F,f}(x) =e^{i\left[\sqrt{\frac 2 3} \phi_+
-2k_F x\right]} ( \Psi^\dagger_{-,\uparrow}
\Psi^\dagger_{+,\downarrow} +\Psi^\dagger_{+,\uparrow}
\Psi^\dagger_{-,\downarrow} ) + \text{H. c.} ,
\end{eqnarray}
\noindent which enables us to find exactly its correlations. We
introduce $u=u_\sigma^*$ to  simplify the notations. The Green's
functions of the fermions read:
\begin{eqnarray}  \label{eq:green-dirac-+}
G_{++}(x,\tau)&=&-\frac m {2\pi u} \frac{\tau+i\frac x u}{\sqrt{\tau^2+\frac{%
x^2}{u^2}}} K_1\left(m\sqrt{\tau^2+\left(\frac x u\right)^2}\right) \\
G_{--}(x,\tau)&=&-\frac m {2\pi u} \frac{\tau-i\frac x u}{\sqrt{\tau^2+\frac{%
x^2}{u^2}}} K_1\left(m\sqrt{\tau^2+\left(\frac x u\right)^2}\right) \\
G_{-+}(x,\tau)&=&G_{+-}(x,\tau)=-\frac m {2\pi u} K_0\left(m\sqrt{%
\tau^2+\left(\frac x u\right)^2}\right)
\end{eqnarray}

Using (\ref{eq:rho}) and  Wick's theorem we find that in real
space the density density correlations read:
\begin{eqnarray}\label{eq:fermion-le-realspace}
\langle T_\tau \rho_{2k_F,f}(x,\tau) \rho_{-2k_F,f}(0.0)\rangle = 2 \left(\frac{m}{%
2\pi u}\right)^2
\left(\frac{\alpha^2}{x^2+(u\tau)^2}\right)^{\frac {K_+}{6}}
\left[K_0^2\left(m\sqrt{\tau^2+\left(\frac x u\right)^2}\right) +
K_1^2\left(m\sqrt{\tau^2+\left(\frac x u\right)^2}\right)\right].
\end{eqnarray}
These correlation functions decay exponentially, with a correlation length $%
u/m=\xi$. Note that expression~(\ref{eq:fermion-le-realspace}) is exact.

On the other side the boson density is given by:
\begin{eqnarray}
\rho_{2k_B,b}(x) =e^{i\left[\sqrt{\frac 8 3} \phi_+ -2k_B x\right]}
e^{i(\Theta_\uparrow +\Theta_\downarrow)} + H. c.
\end{eqnarray}
To calculate the correlation functions in this case, we can use
the equivalence of the Dirac fermions in (1+1)D with the 2D non
critical Ising
model\cite{luther_ising,zuber_77,kadanoff_gaussian_model,schroer_ising,boyanovsky_ising}
to express the boson fields in terms of the order and disorder
parameters of two non-critical Ising models,
$\sigma_{1,2},\mu_{1,2}$ respectively, by:
\begin{eqnarray}
\cos \Theta_\sigma &=& \sigma_{1} \mu_{2} \\ \sin \Theta_\sigma
&=& \mu_{1} \sigma_{2} \\ \cos \Phi_\sigma &=& \sigma_{1}
\sigma_{2} \\ \sin \Phi_\sigma &=& \mu_{1} \mu_{2}
\end{eqnarray}
We find that:
\begin{eqnarray}
\langle T_\tau\rho_{2k_B,b}(x,\tau) \rho_{2k_B,b}(0,0)\rangle \sim \left(%
\frac{\alpha^2}{x^2+(u\tau)^2} \right)^{\frac {2K_+}{3}} 4 \langle
\sigma(x,\tau) \sigma(0,0) \rangle^2 \langle \mu(x,\tau) \mu(0,0)
\rangle^2,
\end{eqnarray}
where we have used   $\langle \sigma_{1,2}(x,\tau)
\sigma_{1,2}(0,0)\rangle=\langle \sigma(x,\tau) \sigma(0,0)\rangle$
and similarly for $\mu$. In the bosonic case, the mapping on the
2D Ising model allows  to calculate the correlation functions
using the results of Ref.~\cite{wu_ising}, where an exact
expression of the correlation functions of the Ising model in
terms of Painlev\'e III transcendants\cite{ince_odes} was derived.
In fact, since we are interested in the low-energy, long-distance
properties of the system, it is enough to replace the Painlev\'e
transcendants with an approximate expression in terms of modified
Bessel functions. The resulting approximate expression is:

\begin{eqnarray}\label{eq:boson-le-realspace}
\langle T_\tau\rho_{2k_B,b}(x,\tau) \rho_{2k_B,b}(0,0)\rangle \sim \left(%
\frac{\alpha^2}{x^2+(u\tau)^2} \right)^{\frac {2K_+}{3}} K_0^2\left(m 
\sqrt{\tau^2+\left(  \frac x u \right)^2}\right).
\end{eqnarray}

Knowing the correlation function in Matsubara space allows us to
obtain them in the reciprocal space via Fourier transforms.
 The Fourier transform of the
density-density response functions (\ref{eq:fermion-le-realspace}) and (\ref{eq:boson-le-realspace})  can be obtained in Matsubara
space from  integrals derived in the Appendix
\ref{app:integral}.

  We find that the bosonic structure factor is:
\begin{eqnarray}  \label{eq:boson_structure_factor}
\chi_{\rho\rho}^B(\pm 2k_B+q,\omega)=\frac {2\pi}{u} \left(\frac{m\alpha}{u}%
\right)^{\frac{4K_+}{3}} \left(\frac m u\right)^2 \frac{\sqrt{\pi}%
\Gamma\left(1-\frac{2K_+}{3}\right)^3}{4\Gamma\left(\frac 3 2 -\frac{2K_+}{3}%
\right)} {}_3F_2\left(1-\frac{2K_+}{3},1-\frac{2K_+}{3},1-\frac{2K_+}{3}%
;\frac 3 2 -\frac{2K_+}{3},1;-\frac{\omega^2+(uq)^2}{4m^2}\right),
\end{eqnarray}

\noindent where $\Gamma(x)$ is the Gamma function and
${}_3F_2(\ldots;\ldots;z)$ is a generalized hypergeometric
function.\cite{erdelyi_functions_1}
Of course, since Eq.~(\ref{eq:boson-le-realspace}) is a long
distance approximation, the expression (\ref{eq:boson_structure_factor}) is also approximate. More
precisely, the exact expression possesses  thresholds at higher frequencies associated with the excitation of more than one pair of kinks in the intermediate
state.  However, the expression~(\ref{eq:boson_structure_factor}) is
\emph{exact} as long as $\omega$ is below the lowest of these
thresholds.
For the fermions, the expression of the structure factor is exact and reads:
\begin{eqnarray}  \label{eq:fermion_structure_factor}
\chi_{\rho\rho}^F(\pm 2k_F + q,\omega)&=&\frac 1 {2\pi u} \left(\frac{m\alpha%
}{u}\right)^{\frac{K_+}{3}} \left[ \frac{\Gamma\left(1-\frac{K_+}{6}\right)^3%
}{\Gamma\left(\frac 3 2 -\frac{K_+}{6}\right)} {}_3F_2\left(1-\frac{K_+}{6}%
,1-\frac{K_+}{6},1-\frac{K_+}{6};\frac 3 2 -\frac{K_+}{6},1;-\frac{%
\omega^2+(uq)^2}{4m^2}\right)\right.  \nonumber \\
&&\left. + \frac{\Gamma\left(2-\frac{K_+}{6}\right)\Gamma\left(1-\frac{K_+}{6%
}\right)\Gamma\left(-\frac{K_+}{6}\right)}{\Gamma\left(\frac 3 2 -\frac{K_+}{%
6}\right)} {}_3F_2\left(2-\frac{K_+}{6},1-\frac{K_+}{6},-\frac{K_+}{6};\frac
3 2 -\frac{K_+}{6},1;-\frac{\omega^2+(uq)^2}{4m^2}\right) \right]
\end{eqnarray}
The response functions are then obtained by the substitution $i\omega\to
\omega+i0$. Since the generalized hypergeometric functions $%
{}_{p+1}F_p(\ldots;\ldots;z)$ are analytic for $|z|<1$ \cite
{slater66_hypergeom_book}, the imaginary part of the response functions is
vanishing for $\omega<2m$. For $\omega>2m$, the behavior of the imaginary
part is obtained from a theorem quoted in \cite
{olsson66_gen_hypergeometric,buehring01_hypergeometric}. According to the
theorem,
\begin{eqnarray}
\frac{\Gamma(a_1)\ldots \Gamma(a_p)}{\Gamma(b_1)\ldots \Gamma(b_p)} {}%
_{p+1}F_p(a_1\ldots a_{p+1};b_1\ldots b_p; z) =\sum_{m=0}^\infty g_m(0)
(1-z)^m + (1-z)^{s_p} \sum_{m=0}^{\infty} g_m(s_p) (1-z)^m,
\end{eqnarray}
provided that:
\begin{eqnarray}
s_p=\sum_{i=1}^p b_i -\sum_{i=1}^{p+1} a_i,
\end{eqnarray}
is not an integer. Therefore, if $s_p<0$, the generalized hypergeometric
function possesses power-law divergence as $z\to 1$. If $0<s_p<1$, it has a
cusp singularity.

\begin{figure}[htbp]
\centering
\includegraphics{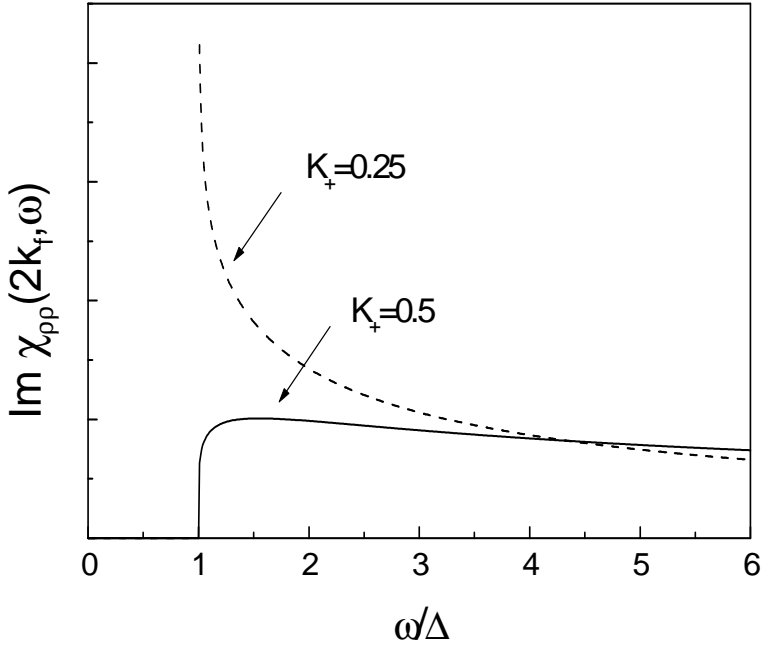}
\caption{The imaginary part of the density-density correlation
function for the fermionic system with $K_+=1/4,1/2$.}
\label{fig:imcorrfun}
\end{figure}

In our case, for the fermion problem, we have
\begin{eqnarray}
s^F_2=1+\frac 3 2 -\frac{K_+}{6} -3 \left(1-\frac{K_+}{6}\right)= \frac{K_+}{%
3}-\frac 1 2
\end{eqnarray}
and in the boson problem
\begin{eqnarray}
s_2^B=1+\frac 3 2 -\frac 2 3 K_+ -3\left(1-\frac 2 3 K_+\right) = \frac{4K_+%
}{3}-\frac 1 2
\end{eqnarray}
Therefore, for $K_+$ small, both the fermion and the boson density-density
response functions show power law singularities for $\omega\to 2m$. For $%
K_+=3/8$, the singularity in the boson density-density response is
replaced by a cusp. This cusp disappears when $K_+=9/8$. For
$K_+=3/2$ the singularity in the fermion density density
correlator also disappears. The behavior of the imaginary part of
the boson density-density correlation function is shown in
Fig.\ref{fig:imcorrfun}.

An exact expression of the imaginary part of the ${}_3F_2$ function can
 be deduced from the calculations in \cite{olsson66_gen_hypergeometric}%
. Indeed, in \cite{olsson66_gen_hypergeometric}, it was found that:
\begin{eqnarray}
{}_3F_2(a_1,a_2,a_3;b_1,b_2;z)=F_R(a_1,a_2,a_3;b_1,b_2;z) + \frac{%
\Gamma(b_1)\Gamma(a_1+a_2+a_3-b_1-b_2)}{\Gamma(a_1)\Gamma(a_2)\Gamma(a_3)}%
\xi(a_1,a_2,a_3;b_1,b_2;z),
\end{eqnarray}
\noindent where $F_R$ is defined by a series that converges absolutely  for $%
\mathrm{Re}(z)>1/2$ and thus has no cut along $[1,+\infty]$ ,  and
$\xi$ is singular along $[1,+\infty]$. $\xi$ can be expressed in
terms of a higher hypergeometric function of two variables, the
Appell function $F_3$, as:
\begin{eqnarray}
\xi(a_1,a_2,a_3;b_1,b_2;z) &=& z^{a_1-b_1-b_2+1} (1-z)^{b_1+b_2-a_1-a_2-a_3}
\nonumber \\
&& \times F_3(b_1-a_1,1-a_2,b_2-a_1,1-a_3,b_1+b_2-a_1-a_2-a_3+1,1-1/z,1-z).
\end{eqnarray}
Since only $\xi$ has a cut along $[1,\infty]$, for $\omega^2>(vq)^2+4m^2$, $%
z>1$, the imaginary part can be expressed as a function of $F_3$
only. This is particularly useful in the case of the fermion
density correlators, because the
expression~(\ref{eq:fermion_structure_factor}) is exact. In the
case of the bosons, thresholds associated with the excitation of a
larger number of Majorana fermions will appear at energies $\sim
4m$. The imaginary parts of correlation functions
Eq.~(\ref{eq:boson_structure_factor})--~(\ref{eq:fermion_structure_factor})
can be measured by Bragg
spectroscopy\cite{stenger99_bragg_bec,stamper-kurn99_bragg_bec,zambelli00_dynamical_structure_factor}.
In Fig.\ref{fig:imcorrfun} we plot the imaginary part of density
correlation functions for the fermionic system with $K_+=1/4$ and $K_+=1/2$ as a function of frequency.
In the first case, we obtain a divergence of the density-density response near the threshold, whereas in the second case, only a cusp is obtained.

\subsubsection{spectral functions of the fermions}

We also wish to calculate the spectral functions of the original fermions $%
\psi_{r,\sigma}$ (not the pseudofermions $\Psi_{r,\sigma}$). To
obtain these spectral functions, we express the operators
$\psi_{r,\sigma}$ as a function of the fields
$\phi_+,\Phi_{\uparrow,\downarrow}$ and their dual fields.

We find:
\begin{eqnarray}
\psi_{+,\sigma}(x)&=&\frac 1 {\sqrt{2\pi\alpha}} e^{\frac i {\sqrt{6}}
(\theta_+-\phi_+)} e^{i \left[-\Theta_{-\sigma} +\frac 5 6 \Phi_{-\sigma}%
\right]} e^{-\frac i 6 \Phi_\sigma} \\
\psi_{-,\sigma}(x)&=&\frac 1 {\sqrt{2\pi\alpha}} e^{\frac i {\sqrt{6}}
(\theta_++\phi_+)} e^{i \left[\Theta_{\sigma} +\frac 5 6 \Phi_{\sigma}\right]%
} e^{-\frac i 6 \Phi_{-\sigma}}
\end{eqnarray}

Therefore, the Green's functions of the original fermions factorize as:
\begin{eqnarray}
-\langle T_\tau \psi_{+,\sigma}(x,\tau) \psi_{+,\sigma}(0,0)\rangle =
G_{+}(x,\tau) G_{-\sigma}^{-1,5/6}(x,\tau) G_{\sigma}^{0,1/6}(x,\tau),
\end{eqnarray}
\noindent where:
\begin{eqnarray}
G_{+}(x,\tau) = -\langle T_\tau e^{\frac i {\sqrt{6}}
(\theta_+-\phi_+)(x,\tau)} e^{-\frac i {\sqrt{6}} (\theta_+-\phi_+)(0,0)}
\rangle=\frac{\alpha}{u\tau -ix} \left(\frac{\alpha^2}{x^2+(u\tau)^2}%
\right)^{\frac 1 {24}\left(\sqrt{K_+}-\frac 1 {\sqrt{K_+}}\right)^2},
\end{eqnarray}

\begin{eqnarray}  \label{eq:for-LZ}
G_{-\sigma}^{-1,5/6}(x,\tau)=-\langle T_\tau e^{i \left[-\Theta_{-\sigma}
+\frac 5 6 \Phi_{-\sigma}\right](x,\tau)} e^{i \left[\Theta_{\sigma} -\frac
5 6 \Phi_{\sigma}\right](0,0)}\rangle,
\end{eqnarray}
and:
\begin{eqnarray}  \label{eq:for-Bernard}
G_{-\sigma}^{0,1/6}(x,\tau)=-\langle T_\tau e^{-\frac i 6
\Phi_\sigma(x,\tau)} e^{\frac i 6 \Phi_\sigma(0,0)} \rangle.
\end{eqnarray}
The correlator in Eq.~(\ref{eq:for-Bernard}) satisfies differential
equations that were derived in \cite{bernard94_ising_equadiff}. However,
since the fields $\Phi_\sigma$ are long range ordered, we can simply replace
the terms $e^{\pm \frac i 6 \Phi_\sigma}$ by their expectation value $%
\langle e^{\pm \frac i 6 \Phi_\sigma}\rangle$. This approximation
only affects the behavior of the fermion correlator at high
energy.

Therefore, we are left with~(\ref{eq:for-LZ}) to evaluate. To do this, we
can use exact results derived in \cite
{lukyanov_soliton_ff,essler_quarter_filled,tsvelik_spectral_cdw} to obtain a
form-factor expansion of Eq.~(\ref{eq:for-LZ}). The first term of the Form
factor expansion yields:
\begin{eqnarray}
G_{\sigma}^{-1,5/6}(x,\tau)=\int \frac{d\psi}{2\pi} e^{\frac 5 6
\psi}
e^{m(i\frac x u \sinh \psi - \tau \cosh \psi)} + O(e^{-3m\sqrt{\tau^2+(x/u)^2%
}})
\end{eqnarray}
Writing:
\begin{eqnarray}
u \tau =\rho \cos \varphi \\
x = \rho \sin \varphi
\end{eqnarray}

We finally obtain that:
\begin{eqnarray}
G(x,\tau)\sim e^{i\varphi} \left(\frac \alpha \rho\right) ^{\frac
1 {12}\left( K_++\frac 1 {K_+}\right)} K_{\frac 5 6} \left(\frac m
u \rho\right).
\end{eqnarray}

The Fourier transform of $G(x,\tau)$ is given by a Weber-Schaefheitlin
integral\cite
{weber_erdelyi,weber_gradshteyn,weber_magnus,tsvelik_spectral_cdw,essler_quarter_filled}%
. In final form, the Fourier transform of the Fermion Green's function
reads:
\begin{eqnarray}
\hat{G}(q,\omega_n)\sim {}_2F_1\left(\frac 7 4 -\frac 1 {24}
(K_++K_+^{-1}),\frac {13} {12} -\frac 1 {24} (K_++K_+^{-1});2;-\frac{%
(uq)^2+\omega_n^2}{m^2}\right)
\end{eqnarray}

When this Green's function is analytically continued to real
frequency, it is  seen\cite{abramowitz_math_functions} that
it has a power law singularity for $\omega^2=(uq)^2+m^2$, and is
analytic for $\omega$ below this threshold. As a result, the
Fermion Green's function vanishes below the gap, as it would do in
a superconductor\cite{abrikosov_book} and could be checked
experimentally. Note however that the anomalous Green's function
remains zero\cite{orignac03_gorkov}, due to the fluctuations of
the phase $\theta_+$.

\section{Mott insulating state}

\label{sec:mott-insul-state} Till now, we have only considered the case of
the continuum system~(\ref{eq:nolattice}) or incommensurate filling in the
lattice system~(\ref{eq:lattice}). We now turn to a lattice system at
commensurate filling. We first establish a generalization of the
Lieb-Schultz-Mattis theorem\cite{lieb_lsm_theorem,affleck_lieb,rojo_lsm_generalized,affleck_lsm,yamanaka_luttinger_thm,cabra_ladders} in
the case of the boson-fermion mixture described by the
Hamiltonian~(\ref{eq:lattice}). This will give us a condition for the
existence of a Mott insulating state without spontaneous breakdown of
translational invariance. Then, we will discuss using bosonization the
properties of the Mott state. We note that Mott states have been
studied in the boson-fermion model in
Refs.~\cite{zhou05_mott_bosefermi,zhou05_mott_bosefermi_long},
 but not in a one-dimensional case. Finally, we will consider the case
when the molecules or the atoms can form a Mott insulating case in the
absence of boson-fermion conversion, and we will show that this Mott state
is unstable.

\subsection{Generalized Lieb-Schultz-Mattis theorem}

\label{sec:gener-lieb-schulz} A generalized Lieb-Schultz-Mattis
theorem can be proven for the boson fermion mixture described by
the lattice
Hamiltonian~(\ref{eq:lattice})\cite{lieb_lsm_theorem,affleck_lieb,affleck_lsm,yamanaka_luttinger_thm}.
Let us introduce the operator:
\begin{eqnarray}
U=\exp\left[i \frac {2\pi}{N} \sum_{j=1}^N (2 b^\dagger_jb_j +
f^\dagger_{j,\downarrow}f_{j,\downarrow} +
f^\dagger_{j,\uparrow}f_{j,\uparrow}) \right],
\end{eqnarray}
such that $U^\dagger H_{bf} U=H_{bf}$. Following
the arguments in Ref.\cite{yamanaka_luttinger_thm}, one has:
\begin{eqnarray}
\langle 0 | U^\dagger H U- H_|0 \rangle &=& O\left(\frac 1 N\right) \\
U^\dagger T U &=& T e^{i 2\pi \nu},
\end{eqnarray}
\noindent where $|0\rangle$ is the ground state of the system, $T$ is  the
translation operator and $H$ is the full Hamiltonian. The  quantity $\nu$ is
defined by:
\begin{eqnarray}
\nu=\frac 1 N \sum_{j=1}^N (2 b^\dagger_j b_j + \sum_\sigma
f^\dagger_{j,\sigma} f_{j,\sigma}) = \frac 1 N (2N_b + N_f)
\end{eqnarray}
For noninteger $\nu$, it results from the analysis of \cite
{yamanaka_luttinger_thm} that there is a state $U|0\rangle$  of momentum $%
2\pi\nu\ne 0 [2\pi]$ which is orthogonal to the ground state
$|0\rangle$ and is only $O(1/N)$ above the ground state. This
implies either a ground state degeneracy (associated with a
spontaneous breaking of translational symmetry) or the existence
of gapless excitations (if the spontaneous translational symmetry
is unbroken and the ground state is unique). For integer $\nu$,
the ground state and the state $U|0\rangle$ have the same
momentum. In that case, a gapped state without degeneracy can be
obtained. This state is analogous to the Mott insulating state in
the half-filled Hubbard model in
one-dimension\cite{lieb_hubbard_exact} or the Mott insulating
state in the Bose-Hubbard model with one boson per
site\cite{kuhner_bosehubbard}. We note that for $\lambda=0$ in the
Hamiltonian (\ref{eq:lattice}) fermions and bosons are separately
conserved, and the corresponding Fermi wavevectors are:
\begin{eqnarray}
k_B&=&\frac {\pi N_b}{N} \\
k_F&=&\frac {\pi N_f}{2N}
\end{eqnarray}
The momentum of the state $U|0\rangle$ is thus equal to $4(k_B +k_F)$. The
condition to have a Mott insulating state in the Hubbard model, $4k_F=2\pi$
is thus generalized to $4(k_B+k_F)=2\pi$ i.e. $2N_b+N_f=N$.

\subsection{umklapp term}
\label{sec:umklapp}

In this Section, we provide a derivation of the umklapp term valid
in the case of the lattice system~(\ref{eq:lattice}). Let us
consider the $2k_F$ and $2k_B$ components of the atom and molecule
charge density, given respectively by
Eq.~(\ref{eq:fermion-density-bosonized}) and
Eq.~(\ref{eq:boson-density-bosonized}). These terms yield an
interaction of the form:
\begin{eqnarray}
  \label{eq:comm-repuls}
  && C \int dx \cos (2\phi_b -2k_B x) \cos (\sqrt{2}\phi_\rho -2k_F x)
  \cos \sqrt{2} \phi_\sigma \nonumber \\
  && = \frac C 2 \int   \cos [2\phi_b +\sqrt{2}\phi_\rho -2(k_B+k_F)
  x]\cos \sqrt{2} \phi_\sigma \nonumber \\
  && + \frac C 2 \int   \cos [2\phi_b -\sqrt{2}\phi_\rho -2(k_B-k_F)
  x]\cos \sqrt{2} \phi_\sigma
\end{eqnarray}

In Eq.~(\ref{eq:comm-repuls}), the last line is the backscattering
term of (\ref{eq:cdw-locking}), and the second line is the umklapp
term. Let us consider a case with $k_F\ne k_B$, and let us
concentrate on the effect of the umklapp term.
 Using
the rotation~(\ref{eq:rot}), we can reexpress it as:
\begin{eqnarray}
\label{eq:umk-source}
= \frac C 2 \int   \cos [\sqrt{6}\phi_+ -2(k_B+k_F)
  x]\cos \sqrt{2} \phi_\sigma \nonumber \\.
\end{eqnarray}
In the following we will consider the cases corresponding to one
or two atoms per site.

\subsubsection{Mott insulating state with one atom per site}

Let us consider first the case of $(k_B+k_F)=\frac \pi {2\alpha}$. Then,
the term (\ref{eq:umk-source}) is oscillating. In second
order perturbation theory, it  gives rise to the umklapp term:
\begin{eqnarray}  \label{eq:umklapp}
H_{umk.}^{1F}=\frac{2g_U}{(2\pi\alpha)^2} \int dx \cos \sqrt{24}\phi_+.
\end{eqnarray}
\noindent The condition for the appearance of the umklapp
term~(\ref{eq:umklapp}) can be seen to correspond to having one
fermion atom per site of the atomic lattice. Let us briefly
mention two alternative derivations of (\ref{eq:umklapp}). A
simple derivation   can be obtained by considering
 the combination of  the $4k_B$ term in
the boson density with the $4k_F$ term in the fermion density in
Haldane's expansion\cite{haldane_bosons}. A second derivation can
be  obtained by considering the effect of a translation by one
lattice parameter on the phases $\phi_\rho$ and $\phi_b$ \cite
{yamanaka_luttinger_thm,oshikawa_plateaus}. The expressions of the
 densities (\ref{eq:fermion-density-bosonized}) and
 (\ref{eq:boson-density-bosonized}) imply that upon a translation by a
 single site $ \phi_\rho \to \phi_\rho -\sqrt{2} k_F \alpha$ and
 $\phi_b \to \phi_b- k_B \alpha$. Therefore, the combination
 $\sqrt{6}\phi_+ = 2 \phi_b
 +\sqrt{2}\phi_\rho$ transforms as :$\sqrt{6}\phi_+\to \sqrt{6}\phi_+
 -2 (k_B + k_F) \alpha$. For $2(k_F+k_B)=\pi/\alpha$, the term $\cos 2
 \sqrt{6}\phi_+$ is invariant upon translation, thus leading again to
 (\ref{eq:umklapp}). The presence of the umklapp term~(\ref
{eq:umklapp}) in  the  Hamiltonian can result in the opening of a
charge gap and the formation of a Mott insulating state. Since the
umklapp term is of dimension  $6K_+$ this implies that a  Mott
insulating state is possible only for $K_+<1/3$ i.e. very strong
repulsion. For free fermions, the Mott transition would occur at
$K_\rho=1$ i.e. for weakly repulsive interaction. Thus, we see
that the Josephson coupling is very effective in destabilizing the
Mott state. In the Mott insulating state,  the superfluid
fluctuations become short ranged. Since CDW fluctuations are also
 suppressed, the system  shows some
analogy with the Haldane gapped phase of spin-1
 chains\cite{haldane_gap} in that it is totally quantum disordered.
 In fact, this analogy can be strengthened by exhibiting
an analog of the VBS (valence
bond solid) order parameter\cite{nijs_dof,kennedy_z2z2_haldane}.
In Haldane gapped chains, this nonlocal order parameter measures
a hidden long range order in the  system associated with the breakdown
 of a hidden discrete symmetry in the system. The equivalent nonlocal order
 parameter for the atom-molecule system is discussed in
 Appendix~\ref{app:non_loc_ord}.

\subsubsection{Mott insulating state with two atoms per site}

Another commensurate filling, where a Mott insulating state is
possible is obtained for $(k_F+k_B)=\pi/\alpha$. This case
corresponds to having one molecule (or two atoms) per site of the
optical lattice. In that case, the term in (\ref{eq:umk-source})
is non-oscillating, and it gives rise to an umklapp term of the
form:
\begin{eqnarray}
  \label{eq:umklapp-boson}
  H_{umk}^{1B}=\frac{2g_U}{(2\pi\alpha)^2} \int dx \cos \sqrt{6}\phi_+ \cos
  \sqrt{2}\phi_\sigma.
\end{eqnarray}
We notice that this umklapp term is compatible with the  spin gap
induced by the Josephson term (\ref{eq:lambda-bosonized}). When
the Josephson coupling is large, we can make $\cos
\sqrt{2}\phi_\sigma \to \langle \cos \sqrt{2}\phi_\sigma\rangle$
and we see that the term (\ref{eq:umklapp-boson}) becomes relevant
for $K_+=4/3$. For weaker Josephson coupling, the dimension
becomes $1/2+3/2 K_+$, and this term is relevant only for $K_+<1$.
Since $K_+=1$ corresponds to hard core bosons, this means that for
weak Josephson coupling, the Mott state with a single boson per
site becomes trivial. Interestingly, we note that increasing the
Josephson coupling is \emph{enhancing} the tendency of the system
to enter a Mott insulating state as a result of the formation of a
spin gap. If we compare with a system of bosons at commensurate
filling, we note however that the Mott transition would obtain for
$K_b=2$\cite{giamarchi_book_1d}. Therefore, the Josephson coupling
still appears to weaken the tendency to form a Mott insulating
state. Such tendency was also observed in
\cite{zhou05_mott_bosefermi}.

\subsection{Commensurate filling of the atomic or molecular
subsystem}

When the atomic subsystem is at commensurate filling
($4k_F=\frac{2\pi}{a}$), an umklapp term:
\begin{equation}
  \label{eq:umklapp-fermions}
  \frac{-2 g_3}{(2\pi\alpha)^2} \cos \sqrt{8} \phi_\rho,
\end{equation}
must be added to the Hamiltonian. Such umklapp term can create a
gap in the density  excitations of the unpaired atoms. However, we
must also take into account the term~(\ref{eq:lambda-bosonized}).
This term is ordering $\theta_-$ and thus competes with the
umklapp term (\ref{eq:umklapp-fermions}).  To understand what
happens when $\theta_-$ is locked, it is convenient to rewrite the
umklapp term (\ref{eq:umklapp-fermions}) as $\propto \cos
\sqrt{8/3} (\sqrt{2} \phi_+ - \phi_-)$. The terms generated by the
renormalization group are of the form $\cos n\sqrt{8/3} (\sqrt{2}
\phi_+ - \phi_-)$, with $n$ an integer. When $\theta_-$ is locked,
replacing the terms $e^{i\beta \phi_-}$ by their expectation
values, we find that all these terms vanish. Therefore, no term
$\cos \beta \phi_+$ can appear in the low energy Hamiltonian. A
more formal justification of the absence of the $\cos \beta
\phi_+$ term in the low energy Hamiltonian can be given by noting
that when the Hamiltonian is expressed in terms of $\phi_{\pm}$ it
has a continuous symmetry $\phi_+ \to \phi_++\alpha$ and $\phi_-
\to \phi_-+\sqrt{2} \alpha$. As a result, terms of the form $\cos
\beta \phi_+$ are forbidden by such symmetry. The consequence of
the absence of $\cos \beta \phi_+$  terms in the Hamiltonian when
$\theta_-$ is locked is that, even if the unpaired atom density is
at a commensurate filling, the umklapp terms do not destabilize
the coupled phase. However, in the opposite case of a strong
umklapp term and a weak boson-fermion conversion term, it is the
field $\phi_\rho$ that will be ordered. The previous arguments can
be reversed and show that the formation of a Mott gap for the
fermions will prevent the formation of the coupled phase. Using
the method of Ref.\cite{jose_planar_2d}, one can show that the
phase transition between the coupled and the decoupled state is
identical to the phase transition that occurs in two
non-equivalent coupled two-dimensional XY models. This phase
transition was studied by the renormalization group in
\cite{nelson80_smectics,granato86_xy_coupled}. It was found that
in the case of interest to us, this phase transition was in the
Ising universality class. Thus, one expects a quantum Ising phase
transition between the state where the fermions are decoupled from
the bosons and form a Mott insulator and the state where the
fermions and bosons are coupled and form a superfluid.

Of course, the same arguments can also be applied to the bosons
at commensurate filling, the role of the fields $\phi_b$ and $\phi_\rho$
being simply reversed.

\section{Relation with experiments}\label{sec:param-boson-hamilt}
\subsection{Without a potential along the tubes}\label{sec:expt-no-pot}

To connect experiments in quasi-one-dimensional confining
waveguides with theoretical models in 1D, it is necessary to
obtain estimates of the parameters that enter the Hamiltonians
(\ref{eq:lattice}),(\ref{eq:nolattice}) and the bosonized
Hamiltonian (\ref{eq:bosonized-spin}),(\ref{eq:lambda-bosonized})
and  (\ref{eq:detun-bosonized}). Since the parameters in the
Hamiltonian (\ref{eq:lattice}) depend on the periodic optical
trapping potential, we will mainly focus on the parameters that
enter in the continuum Hamiltonian Eq.~(\ref{eq:nolattice}) and in
the bosonized Hamiltonian, i.e. the Luttinger exponent $K_\rho$,
the velocity, and the fermion-boson coupling $\lambda$. Before
giving an estimate of the parameters we need first to remind that,
at the two-body level, there is a connection between the 1D
boson-fermion model and the quasi-1D single channel
model\cite{fuchs04_resonance_bf}, thus we will use one or the
other depending on the physical parameter wherein we are
interested. Experimentally, molecules have been formed from
fermionic atoms
${}^6$Li\cite{zwierlein_bec,jochim_bec,strecker_bec,cubizolles_bec}
and ${}^{40}$K\cite{greiner_bec,regal_bec,moritz05_molecules1d}.
For ${}^6$Li, the mass is roughly 6 times the mass of the proton,
$m_F({}^6\mathrm{Li})=9.6\times 10^{-27}\mathrm{kg}$, and for
${}^{40}$K it is $m_F({}^{40}K)=6.4\times 10^{-26}\mathrm{kg}$.
Then, we need  to determine the effective interaction of atoms
under cylindrical confinement. For the interaction, we will assume
a contact form, i.e. $V(x-x')= g_{1D}\delta(x-x')$ and the
Hamiltonian describing the confined atoms reads:
\begin{eqnarray}
\label{eq:atatham} H&=&-\frac{\hbar^2}{2m} \int d \mathbf{r}
\psi^\dagger_\sigma (\mathbf{r} ,t) \triangle \psi_\sigma
(\mathbf{r},t)+ \frac{m}{2} \int d \mathbf{r} \psi^\dagger_\sigma
(\mathbf{r},t) (\omega_\perp^2\mathbf{r}_\perp^2 +\omega_z z^2)
\psi_\sigma (\mathbf{r},t) \nonumber \\ && + \frac 1 2 \int d
\mathbf{r} d \mathbf{r'} \psi^\dagger_\sigma
(\mathbf{r},t)\psi^\dagger_\sigma (\mathbf{r'},t)
U(\mathbf{r}-\mathbf{r'})
\psi_\sigma(\mathbf{r'},t)\psi_\sigma(\mathbf{r},t),
\end{eqnarray}
where  $\mathbf{r}=(\mathbf{r_\perp},z)$, the second term
represents the harmonic confinement potential, and:
\begin{equation}\label{eq:3dcontact}
U(\mathbf{r})= g_{3D} \delta (\mathbf{r}),
\end{equation}
is the atom-atom repulsion. The coupling  constant $g_{3D}$ is
expressed as a function of the atom-atom scattering length $a_s$
as\cite{abrikosov_book}
\begin{eqnarray}\label{eq:g-scatt-length}
  g_{3D}= \frac{4 \pi \hbar^2 a_s}{m}.
\end{eqnarray}
We introduce the following decomposition of the fermion
annihilation operator:
\begin{equation}
\psi (\mathbf{r},t)=\sum_m \phi_m(\mathbf{r_\perp})
\psi_{m\sigma}(z,t),
\end{equation}
where
$\{\psi_{m\sigma}(z),\psi_{m'\sigma'}(z')\}=\delta_{mm'}\delta_{\sigma,\sigma'}\delta(z-z')$,
and the eigenstates of the transverse Hamiltonian $\phi_m$ satisfy
the Schr\"odinger equation:
\begin{equation}
\left( -\frac{\hbar^2}{2m} \triangle_\perp +
\frac{m\omega_\perp^2}{2} \mathbf{r}_\perp^2\right) \phi_n
(\mathbf{r}_\perp)=\hbar \omega_0 (n +\frac  1 2 )  \phi_n
(\mathbf{r}_\perp),
\end{equation}
and the orthonormal condition $\int  d\mathbf{r}_\perp \phi_n
(\mathbf{r}_\perp) \phi_m (\mathbf{r}_\perp)=\delta_{n,m}$. We
will assume that we have a very elongated trap, with $\omega_z\ll
\omega_\perp$ so that we can neglect the longitudinal confinement.
The first line of the Hamiltonian~(\ref{eq:atatham}) can then be
rewritten as:
\begin{eqnarray}
  H_0=\sum_n \int dz \left[ -\frac{\hbar^2}{2m}
  \psi_{n,\sigma}^\dagger(z,t) \triangle \psi_{n,\sigma}(z,t) + \hbar
  \omega_0(n+1/2) \psi_{n,\sigma}^\dagger(z,t)  \psi_{n,\sigma}(z,t)
  \right]
\end{eqnarray}

If  the transverse zero-point energy is much higher than the
interaction energy per atom,  the transverse motion is frozen in
the ground state. It is known that virtual transitions to higher
states can give rise to a divergence in 1D scattering length
 known as confinement induced resonance (CIR) which is a kind of Fano-Feshbach
resonance\cite{olshanii_cir,bergeman_cir,yurovsky_feshbach,astrakharchik_bose_1d}.
We will ignore CIR
for the moment and  restrict ourselves to consider the
lowest energy level $n=0$ for the transverse motion.
The interaction in (\ref{eq:atatham}) can be rewritten as:
\begin{equation}
H_{int}=\frac  1 2 \int dz \int dz'
\psi^\dagger_{0,\sigma}(z)\psi^\dagger_{0,\sigma}(z')V(z-z')\psi_{0,\sigma}(z')\psi_{0,\sigma}(z),
\end{equation}
where the effective potential $V$ reads:
\begin{equation}
\label{eq:ueff} V(z-z')=\int d\mathbf{r}_\perp d\mathbf{r'}_\perp
|\phi_0(\mathbf{r}_\perp)|^2 |\phi_0(\mathbf{r'}_\perp)|^2
U(\mathbf{r}_\perp-\mathbf{r'}_\perp,z-z'),
\end{equation}
Using the expression of the ground state wave function  of the
transverse motion:
\begin{equation}
\phi_0(\mathbf{r}_\perp)=\sqrt{\frac{m\omega_\perp}{\pi\hbar}}
e^{-\frac{m \omega_\perp}{2\hbar}\mathbf{r}_\perp^2},
\end{equation}
substituting it into (\ref{eq:ueff}), using the definition of the
interaction~(\ref{eq:3dcontact}),  and integrating over the transverse
coordinate, we obtain
$V(z)=g_{1D}\delta(z)$ with the effective one-dimensional
coupling\cite{dunjko_bosons1d,astrakharchik_bose_1d}:
\begin{equation}
\label{eq:effcoup} g_{1D}=\frac{g_{3D}}{2\pi} \frac{m
\omega_\perp}{\hbar}=2 \hbar a_s \omega_\perp.
\end{equation}

Eq.~(\ref{eq:effcoup}) can be generalized to the case in which
atoms of different species, with different trapping frequencies
$\omega_{\perp,1}$ and $\omega_{\perp,2}$ are interacting with each
other. In that case,
\begin{equation}
\label{eq:effcoup-12} g_{1D}=4 \hbar a_{12} \frac{\omega_{\perp
1}\omega_{\perp 2}}{(\omega_{\perp 1}+\omega_{\perp 2})},
\end{equation}
where $a_{12}$ is the atom-atom scattering length. Knowing
$g_{1D}$, we can obtain the Luttinger exponent of fermions in
(\ref{eq:bosonized-spin}) as:
\begin{eqnarray}
  K_\rho=\left(1+\frac{g_{1D}}{\pi \hbar v_F}\right)^{-1/2}.
\end{eqnarray}
In the case of bosons, the Luttinger exponent $K_b$ must be extracted from
the Lieb-Liniger
equations\cite{lieb_bosons_1D,takahashi_tba_review,giamarchi_book_1d}.

Having obtained the form of the effective interactions, we turn to
the determination of  the Josephson coupling $\lambda$ in the boson-fermion
model~(\ref{eq:nolattice}). In the
3D case\cite{duine_feshbach_review,romans04_bose_ising}, the
boson-fermion conversion factor is given by:
\begin{eqnarray}
  \label{eq:BF-3D}
  \lambda_{3D}\int d^3\mathbf{r} \psi^\dagger_B(\mathbf{r})
  \psi_\uparrow(\mathbf{r}) \psi_\downarrow(\mathbf{r}),
\end{eqnarray}
with:
\begin{eqnarray}
  \lambda_{3D}=\hbar \sqrt{\frac{4\pi a_{bg} \Delta \mu \Delta B}{m}},
\end{eqnarray}
\noindent where $a_{bg}$ is the atom-atom scattering length far
from resonance, $\Delta B$ is the width of the resonance and
$\Delta\mu$ is the difference of magnetic moment between atom and
molecule. Using the projection on the lowest level, we obtain:
\begin{eqnarray}
  \lambda^2 =2\hbar \omega_\perp a_{bg} \Delta \mu \Delta B.
\end{eqnarray}

Knowing the interaction $\lambda$ in terms of the relevant
physical parameters, we finally have to determine the spatial
cutoff $\alpha$ to use in the bosonization and comment on the
validity of bosonization approach. In the case of the optical
lattice Eq.~(\ref{eq:lattice}), the obvious spatial cutoff is the
lattice spacing.  The cutoff to use in bosonization for the
continuum case of Eq.~(\ref{eq:nolattice}) is obtained in the
following way. Bosonization is applicable as long as the
 kinetic energy of longitudinal motion of the
particles is much smaller than the trapping energy $\hbar
\omega_\perp$. Thus, we
have to impose the condition: $\hbar v_F \Lambda \sim \hbar
\omega_\perp$, where $\Lambda$ is the momentum
cutoff\cite{dunjko_bosons1d}.
The real
space cutoff in the continuum case
is thus $\alpha \sim \Lambda^{-1}\sim v_F/\omega_\perp$.

The condition for perturbation theory to be valid is that the
energy associated with the formation of molecules, $\lambda
\alpha^{-1/2}$ is small with respect to the energy cutoff $\hbar \omega_\perp$.
Therefore, perturbation theory is applicable when:
\begin{eqnarray}
  \frac{\lambda}{\hbar (v_F \omega_\perp)^{1/2}} \ll 1,
\end{eqnarray}
i.e.
\begin{eqnarray}
  \frac{a_{bg} \mu \Delta B}{\hbar v_F} \ll 1
\end{eqnarray}

Using the values given in Refs.~\cite{strecker_bec,bruun05_6li_interaction},  we
find that this parameter is small for $v_F \gg 3.2\times
10^{-2}$m/s. Since $v_F$ can be expected to be of the order of
$10^{-3}$m/s, this is not unreasonable. In fact, using the values
of the trapping frequency given by Moritz et
al.\cite{moritz05_molecules1d}  we find that:
\begin{eqnarray}
  v_F&=&\sqrt{\frac{2N\hbar \omega_z}{m}} \\
     &=&\sqrt{\frac{2\times 100 {\text m^{-3}}\times 10^{-34} {\text J}\cdot {\text s}\times 10^3 {\text Hz}}{6\times
     1.6\times 10^{-27}{\text Kg}}} \\
     &=& 4.6 \times 10^{-2} \mathrm{m/s}
\end{eqnarray}

Therefore, we see that with ${}^6$Li at the narrow resonance,
the ratio is of order $0.7$
and we can expect that our theory is valid qualitatively.

Concerning $K_\rho$, we find that:
\begin{eqnarray}
  K_\rho \sim 1-\frac{\omega_\perp a_{bg}}{v_F} &\simeq& 1-\frac{2\pi
  \times (69 \times 10^3) s^{-1} \times (80\times 0.5\times 10^{-10}) m}{4.6\times 10^{-2} m.s^{-1}},\\
 &\simeq& 0.995
\end{eqnarray}
i.e. interactions between fermions can be neglected. Since the
interaction between the molecules \cite{petrov05_dimers_scattering}
has a scattering length  $a_{BB}=0.6a_{FF}$ one sees that molecules are
only weakly interacting.

\subsection{With a potential along the tubes}\label{sec:expt-w-pot}

As we have seen in Sec.~\ref{sec:expt-no-pot}, in the case of a 
two-dimensional  optical lattice without periodic potential along the tubes, the repulsion
between the bosonic molecules is weak, making the decoupling transition 
or the Mott transition impossible to observe. 
To increase the effect of the repulsion, one needs 
to increase the effective mass of the atoms by adding a periodic potential 
along the tubes. 
A periodic potential can be imposed along the tubes by placing the atoms in a
three dimensional optical lattice. The atoms experience a potential:
\begin{eqnarray}
  \label{eq:3d-potential}
  V(x,y,z)= V_x \sin^2\left(\frac {2\pi x}{\lambda_l}\right)+  V_y
  \sin^2\left(\frac {2\pi y}{\lambda_l}\right) +  V_z \sin^2\left(\frac
    {2\pi z}{\lambda_l}\right),
\end{eqnarray}
where $\lambda_l$ is the wavelength of the laser radiation, and
$V_x\ll V_y,V_z$ so that the system remains quasi one-dimensional. 
The strength of the potential is measured in unit of the
recoil energy
$E_R=\frac{\hbar^2}{2m}\left(\frac{2\pi}{\lambda_l}^2\right)$ as
$V_x=sE_R$.  Typical values for $s$ are in the range $5$ to $25$.
For lithium atoms\cite{anderson96_li_lattice}, the typical value 
of $E_R$ is $76$kHz. If the
potential is sufficiently strong, the atoms tend to localize in
the lowest trap states near the minima of this potential. In our
case, since the periodic potential along the tubes has shallower
minima than in the transverse directions, the small overlap
between the trap states in the longitudinal direction yields the
single band
Hamiltonian~(\ref{eq:lattice})\cite{jaksch05_coldatoms,dickerscheid_feshbach_lattice}.
An expression of the parameters of the lattice model
(\ref{eq:lattice}) in terms of the microscopic parameters has been
derived in\cite{dickerscheid_feshbach_lattice}. On the lattice,
the Fermi velocity of the atoms and the pseudo-Fermi velocity of
the molecules can be reduced by increasing the depth of the
periodic potential in the longitudinal direction. This allows in
principle to move the system near the decoupling
transition\cite{sheehy_feshbach} or the Mott transition, by reducing $K_\rho$.

A second possible setup\cite{albus03_bosefermi}
 is to use a cigar shaped potential:
\begin{eqnarray}
  \label{eq:cigar-shape}
  V_{\text{cigar}}(x,y,z)=\frac 1 2 m \omega_0^2 (x^2 + \mu^2 \mathbf{r}_\perp^2),
\end{eqnarray}
\noindent with $\mu \gg 1$, so that the atoms and the molecules
are strongly confined in the transverse direction, and to apply a
periodic potential:
\begin{eqnarray}
  V_{\text{periodic}}=V_0 \sin^2 \left(\frac{\pi x} d\right),
\end{eqnarray}
in order to form the one dimensional structure described by the
model~(\ref{eq:lattice}).

The main difficulty of experiments in optical lattices is that the 
reduction of the bandwidth results in a reduction of the Fermi velocity $v_F$. 
Since the perturbative regime is defined by $\lambda \alpha^{1/2}\ll v_F$, this
implies that by increasing the depth of the potential in the longitudinal direction one is also pushing the system in the regime where 
the boson-fermion conversion term must be treated 
non-perturbatively\cite{fuchs04_resonance_bf}. However, in that regime 
there isn't anymore coexistence of atoms and molecules and the decoupling 
transition does not exist. Moreover, in that regime, 
the Mott transition becomes the usual
purely fermionic or purely bosonic Mott transition.\cite{giamarchi_book_1d}

\section{Conclusions}

We have studied a one-dimensional version of the boson-fermion
model using the bosonization technique.  We have found that at
low-energy the system is described by two Josephson coupled
Luttinger liquids corresponding to the paired atomic and molecular
superfluids. Due to the relevance of the Josephson coupling for
not too strong repulsion, the order parameters for the Bose
condensation and fermion superfluidity become identical, while a
spin gap and a gap against the formation of phase slips are
formed. As a result of these gaps, we have found that the charge
density wave correlations decay exponentially, differently from
the phases where only bosons or only fermions are
present\cite{fuchs_bcs_bec,fuchs04_resonance_bf}. We have
discussed the application of a magnetic field that results in a
loss of coherence between the bosons and the fermion and the
disappearance of the gap, while changing the detuning has no
effect on the existence of the gaps
 until either the fermion or the
boson density is reduced to zero. We have discussed the effect of
a backscattering term which induces mutual locking of the density
of bosons and fermions favoring charge density wave fluctuations
resulting in a quantum Ising phase transition between the density
wave phase and the superfluid phase. We have found a Luther-Emery
point where the phase slips and the spin excitations can be
described in terms of pseudofermions. For this special point in
the parameter space, we have derived closed form expressions of
the density-density correlations and the spectral functions. The
spectral functions of the fermions are gapped, whereas the
spectral functions of the bosons remain gapless but with an
enhanced divergence for momentum close to zero.  Finally, we have
discussed the formation of a Mott insulating state in a periodic
potential at commensurate filling. We have first established a
generalization of the Lieb-Schulz-Mattis theorem, giving the
condition for the existence of a Mott-insulating state without
spontaneous breakdown of translational invariance. Then, we have
discussed the properties of the Mott-state in the case of one atom
or two atoms per site showing that in the first case the Josephson
coupling is very effective in destabilizing the Mott state.
Finally, we have considered the case when the atoms or the
molecules can form a Mott state in absence of boson-fermion
conversion and shown that this Mott-state is unstable. To connect
our results with experiments in quasi-one-dimensional confining
waveguides we have derived estimates of the parameters that enter
the bosonized Hamiltonian, as the Luttinger exponents, using the
values of the trapping frequency and density used in experiments.
We have seen that bosons are only weakly interacting and the
necessary small fermionic Luttinger parameter required to realize
a strongly interacting system, render the Mott insulating and
decoupled phases difficult to observe in experiments.  A
nontrivial challenge is the experimental realization of the
coupled Luttinger liquids phase with parameters tunable through
the exactly solvable point (the Luther-Emery point). We suggest
that a Fano-Feshbach resonantly interacting atomic gas confined in
a highly anisotropic (1d) trap and subject to a periodic optical
potential is a promising candidate for an experimental measurement
of the physical quantities (correlation functions) discussed here.
Finally we would like to comment on the fact that an interesting
edge states physics is expected when open boundary conditions  (or
a cut one-dimensional boson-fermion system) are considered. The
existence of edge states at the end of the system could lead to
significant contribution to the density profile that could be
tested in experiments. The physics of the edge states will be
similar to the one of Haldane gap systems, like the valence bond
solid model, and a study along this direction is in progress.
\appendix

\section{Calculation of the integrals in
  Eqs.~(\ref{eq:boson_structure_factor})--~(\ref{eq:fermion_structure_factor})}
\label{app:integral}

In this appendix we will derive a slightly more general integral than
those of Eqs.~(\ref{eq:boson_structure_factor}) and
(\ref{eq:fermion_structure_factor}). Namely, we will consider:
\begin{eqnarray}\label{eq:integral-general}
g(y)= \int_0^\infty K_\mu(u) K_\nu(u)  J_\lambda(y u) u^\alpha du.
\end{eqnarray}
\noindent To find  (\ref{eq:integral-general}) explicitly,
we use the series expansion of the Bessel function $J_\lambda$ from \cite
{abramowitz_math_functions} [Eq. (9.1.10)]. We find that
\begin{eqnarray}
g(y)=\left(\frac y 2 \right)^\lambda \sum_{k=0}^\infty \left( -\frac{y^{2}}{4} \right)^k \frac 1
{\Gamma(k+1) \Gamma(k+\lambda+1)}
\int_0^\infty K_\mu(u) K_\nu(u) u^{2k+\lambda+\alpha} du.
\end{eqnarray}
\noindent The integral that appears in the expansion in powers of $y^2$ is a
well known Weber-Schaefheitlin  integral\cite
{weber_magnus,weber_erdelyi,weber_gradshteyn} with two  modified Bessel
functions. Its expression is:
\begin{eqnarray}
\int_0^\infty K_\mu(u) K_\nu(u) u^{2k+\alpha+\lambda} du &=&
  \frac{2^{2(k-1)+\alpha+\lambda}}{\Gamma\left(2k+\alpha+\lambda+1\right)}
  \Gamma\left(k
  +\frac{1+\nu+\mu+\alpha +\lambda}{2}\right)\Gamma\left(k
  +\frac{1+\nu-\mu+\alpha +\lambda}{2}\right) \nonumber \\ && \times \Gamma\left(k
  +\frac{1-\nu+\mu+\alpha +\lambda}{2}\right) \Gamma\left(k
  +\frac{1-\nu-\mu+\alpha +\lambda}{2}\right) ,
\end{eqnarray}

The resulting expression of $g(y)$  can be rearranged using the duplication formula for the
Gamma function, Eq. (6.1.18) in \cite{abramowitz_math_functions}. The final
expression of $g$ is:
\begin{eqnarray}
g(y)=\left(\frac y 2 \right)^\lambda \frac{\pi^{1/2}}{4} \sum_{k=0}^\infty \frac{\Gamma\left(k
  +\frac{1+\nu+\mu+\alpha +\lambda}{2}\right)\Gamma\left(k
  +\frac{1+\nu-\mu+\alpha +\lambda}{2}\right) \times \Gamma\left(k
  +\frac{1-\nu+\mu+\alpha +\lambda}{2}\right) \Gamma\left(k
  +\frac{1-\nu-\mu+\alpha +\lambda}{2}\right)}{\Gamma\left(k+1+\frac
  {\alpha+\lambda}  2\right)\Gamma\left(k+\frac
  {\alpha+\lambda+1}  2\right) \Gamma(k+\lambda+1) } \frac 1
{k!} \left(-\frac {y^2}{4}\right)^k.
\end{eqnarray}
This series expansion is readily identified with the definition of the
generalized hypergeometric function ${}_4F_3$ given in \cite
{slater66_hypergeom_book}. So we find finally that:

\begin{eqnarray}
&&g(y)=\frac{\sqrt{\pi}}{4} \left(\frac y 2 \right)^\lambda  \frac {\Gamma\left(\frac{1+\nu+\mu+\alpha +\lambda}{2}\right) \Gamma\left(\frac{1+\nu-\mu+\alpha +\lambda}{2}\right)  \Gamma\left(\frac{1-\nu+\mu+\alpha +\lambda}{2}\right) \Gamma\left(\frac{1-\nu-\mu+\alpha +\lambda}{2}\right)}
{\Gamma\left(1+\frac {\alpha+\lambda}  2\right) \Gamma\left(\frac{\alpha+\lambda+1}  2\right) \Gamma(\lambda+1)}  \\
 && \times
  {}_4F_3\left(\frac{1+\alpha+\lambda+\nu+\mu}{2},\frac{1+\alpha+\lambda+\nu-\mu}{2},\frac{1+\alpha+\lambda-\nu+\mu}{2},\frac{1+\alpha+\lambda-\nu-\mu}{2};1+\frac
  {\alpha+\lambda} 2,\frac {\alpha+\lambda+1} 2,1+\lambda; -\frac
  {y^2} 4\right)\nonumber
\end{eqnarray}

For $\nu=\mu$, the function ${}_4F_3$ reduces to a simpler ${}_3F_2$
function. This leads to Eqs.~(\ref{eq:boson_structure_factor}) and
(\ref{eq:fermion_structure_factor}).

\section{Non local order parameter for the Mott state}
\label{app:non_loc_ord}

The Mott insulating state can be characterized by the expectation value of
a non-local order parameter as the Haldane gap state in a spin-1 chain\cite{kennedy_z2z2_haldane,nijs_dof}.
The  non-local order parameter is defined as follows:
\begin{eqnarray}  \label{eq:VBS}
O(k,l)=\langle b^\dagger_k b_k \prod_{j>k}^{l} e^{-i\frac{2\pi}{3}(2
b^\dagger_j b_j + \sum_\sigma f^\dagger_{j,\sigma} f_{j,\sigma})}
b^\dagger_l b_l \rangle
\end{eqnarray}

The string operator:
\begin{eqnarray}  \label{eq:string}
O_{\text{string}}(k,l)= \prod_{j>k}^{l} e^{-i\frac{2\pi}{3}(
2 b^\dagger_j b_j + \sum_\sigma f^\dagger_{j,\sigma} f_{j,\sigma})}
\end{eqnarray}
is a product of exponentials.
As a result of its definition, we see that it is
 counting the number of fermions  located between the sites $k$
and $l$, either unbound or forming a molecule.  To derive a bosonized expression of this operator,  we  notice that $exp(2i\pi b^\dagger_j
b_j)=1$ since $b^\dagger_j b_j$ has only integer eigenvalues and
rewrite the string operator as:
\begin{eqnarray}  \label{eq:string-bis}
\prod_{j>k}^{l} e^{i\frac{2\pi}{3} ( b^\dagger_j b_j -
\sum_\sigma f^\dagger_{j,\sigma} f_{j,\sigma})}
\end{eqnarray}

Using bosonization and Eq.~(\ref{eq:string}), we find:
\begin{eqnarray}  \label{eq:string-bosonized}
O_{\text{string}}(x,x^{\prime})= \exp \left[ i\frac{2\pi}{3}
(\rho_B-\rho_F) (x-x^{\prime}) -\frac{2}{\sqrt{3}}
(\phi_-(x)-\phi_-(x^{\prime})) \right]
\end{eqnarray}

Using (\ref{eq:boson-staggered-gap}) and (\ref{eq:string-bosonized})
we obtain the nonlocal order parameter (\ref{eq:VBS}) as:
\begin{eqnarray}  \label{eq:vbs-bosonized}
O(x,x^{\prime})=\langle e^{i\sqrt{\frac {8}{3}} (\phi_+(x)-\phi_+(x^{%
\prime})) -i\frac{2\pi}{3} (2\rho_B
+\rho_F)(x-x^{\prime})}\rangle.
\end{eqnarray}
In the Mott insulating state with one fermion per site, we have
$4(k_F+k_B)=2\pi(2\rho_B+\rho_F)=2n \pi $ where $n$ is an integer. Taking $x,x'\to \infty$, we see that the expectation value of the order parameter is non-vanishing in the Mott state.

A related VBS type order parameter can be defined as:
\begin{eqnarray}
O^{\prime}(k,l)=\langle \left(\sum_\sigma f^\dagger_{k,\sigma}
f_{k,\sigma}\right) \prod_{j>k}^{l} e^{i\frac{2\pi}{3}(
b^\dagger_j b_j - \sum_\sigma f^\dagger_{j,\sigma}
f_{j,\sigma})}\left(\sum_\sigma f^\dagger_{l,\sigma}
f_{l,\sigma}\right) \rangle
\end{eqnarray}

In bosonized form, we have:
\begin{eqnarray}
O^{\prime}(x,x^{\prime})=\langle e^{i\sqrt{\frac {4}{3}} (\phi_+(x)-%
\phi_+(x^{\prime})) +i\frac{\pi}{3} (2\rho_B
+\rho_F)(x-x^{\prime})}\rangle,
\end{eqnarray}
and again this order parameter is non-vanishing.
The physical interpretation of the non-zero expectation value of
these nonlocal order parameters is that
both bosons and fermions possess a hidden charge density wave
order in the Mott insulator. This charge density wave is hidden as a result
 of the fluctuation of the density of fermions and the density of bosons.


\end{document}